\documentclass[a4paper,11pt]{article}
\usepackage{jheppub} 
\usepackage{lineno}
\usepackage[normalem]{ulem}
\usepackage{bm}

\usepackage{tikz}
\usetikzlibrary{decorations}
\usetikzlibrary{arrows.meta, positioning, shapes.geometric, arrows, calc, decorations.pathreplacing, decorations.markings, decorations.pathmorphing}
\tikzstyle{decision} = [diamond, minimum width=3cm, minimum height=1cm, text centered, draw=black, fill=green!30]

\title{Azimuthal momentum isotropization in the Quark-Gluon Plasma thermalization}

\author[a]{Sergio Barrera Cabodevila,}
\author[b,c,d]{Xiaojian Du,}
\author[e,f]{Carlos A. Salgado,}
\author[e]{and Bin Wu}
\affiliation[a]{Institute for Theoretical Physics, University of Heidelberg, Philosophenweg 16, 69120 Heidelberg, Germany}
\affiliation[b]{Wilczek Quantum Center, Shanghai Institute for Advanced Studies,
University of Science and Technology of China, Shanghai 201315, China}
\affiliation[c]{Shanghai Research Center for Quantum Sciences, Shanghai 201315, China}
\affiliation[d]{Hefei National Laboratory, Hefei 230088, China}
\affiliation[e]{Instituto Galego de Física de Altas Enerxías IGFAE, Universidade de Santiago de Compostela, E-15782 Galicia-Spain}
\affiliation[f]{Axencia Galega de Innovaci\'on (GAIN), Xunta de Galicia, Galicia, Spain}

\emailAdd{barrera@thphys.uni-heidelberg.de}

\abstract{Azimuthal anisotropies coming from the initial state of a heavy-ion collision have been historically disregarded in the study of thermalization because they are expected to be rapidly washed out due to final-state interactions. However, they may be important when one attempts to describe azimuthal correlations observed in the collisions of small systems. In this work, we study how these initial anisotropies relax in the context of the Boltzmann Equation in Diffusion Approximation (BEDA). We find a clear hierarchy in the relaxation time of the anisotropies in terms of each harmonic coefficient. We also explore the evolution of the $p_T$-dependent harmonic coefficients in time, finding a shift in the initial peak towards higher momenta that mimics the experimental data when we perform a phenomenologically motivated simulation.}

\begin{document}

\maketitle
\flushbottom

\section{Introduction}\label{sec:Intro}

The fascinating phenomena of the Quark-Gluon Plasma (QGP) in the laboratory as a product of heavy-ion collisions has resulted in a very active topic of research during the last few decades. Many related open questions remain unsolved to this day, a lot of them linked to the early stages of this collision, when the system has not yet had sufficient time to fully thermalize~\cite{Busza:2018rrf, Schlichting:2019abc, Berges:2020fwq}. These out-of-equilibrium stages do not have any strong experimental evidence. However, a solid theoretical foundation has allowed the scientific community to build up a consistent physical picture of thermalization. In the case of heavy-ion collisions, it is assumed that a large number of gluons are produced in the saturation fashion~\cite{Jalilian-Marian:1996mkd, Kovchegov:1998bi, Mueller:1999fp}. This system, called \textit{glasma}~\cite{Lappi:2006fp}, is quickly expanding and far away from equilibrium, and it can be studied in the Color Glass Condensate (CGC) effective theory~\cite{Gelis:2010nm}, which describes the system in terms of strong fields. The rapid thermalization expected from experimental results can not be explained within this framework, and thus, another regime for the description of the system out of equilibrium must emerge.

This new regime shows up by the time that the fields are not strong enough to be described in the CGC~\cite{Mueller:1999pi, Epelbaum:2014yja}. In this case,  kinetic theory can be employed to explain thermalization. This framework is valid for small coupling and describes how the system equilibrates in the so-called "bottom-up thermalization"~\cite{Baier:2000sb}. This analysis confirms the formation of a thermal state starting from the overpopulated system produced right after the collision. Besides, there is a nice match between the classical statistical simulation for the early times and the dynamics predicted in the kinetic description in terms of the non-thermal attractor~\cite{Berges:2013eia, Berges:2013fga}.

The study of thermalization within the kinetic theory framework has been mostly focused on the full QCD leading-order description provided by the so-called Effective Kinetic Theory (EKT)~\cite{Arnold:2002zm}. In this scenario, multiple studies have explored the hydrodynamization and chemical equilibration~\cite{Kurkela:2015qoa, Kurkela:2018oqw, Kurkela:2018xxd, Du:2020dvp, Du:2020zqg}, finding consistent results with phenomenological estimates. Furthermore, the EKT also offers an interesting opportunity for the study of the properties of hard probes propagating in a medium out of equilibrium or in equilibrium, such as jets~\cite{Boguslavski:2024ezg, Zhou:2024ysb, Boguslavski:2025ylx} and heavy quarks~\cite{Boguslavski:2023fdm, Du:2023izb}. Together with the exploration of the effects in the glasma~\cite{Avramescu:2023qvv, Avramescu:2026qro, Barata:2024xwy, Sun:2019fud, Pandey:2023dzz}, this can be used to produce a complete picture of a hard probe evolution during the initial stages of heavy-ion collision.

The Boltzmann Equation in Diffusion Approximation (BEDA) has been used as a simplified version of the EKT that preserves the relevant physics during thermalization~\cite{BarreraCabodevila:2022jhi, Cabodevila:2023htm, BarreraCabodevila:2025vir}. It drops out large-angle interactions while the small-angle approximation is employed to simplify the $2\leftrightarrow2$ interactions~\cite{Mueller:1999pi}, as they are the ones that allow to obtain the dynamics corresponding to the non-thermal attractor~\cite{Berges:2013fga} and the large-angle interactions are not parametrically more important. Regarding the $1\leftrightarrow2$ splittings and mergings, only the Landau-Pomeranchuk-Migdal (LPM) effect is included and the Bethe-Heitler is disregarded as the latter only dominates at very early times. In the present work, we present a study of the azimuthal isotropization in momentum space fully based on the BEDA framework. 

One of the most relevant experimental signatures of the production of the QGP is the presence of long-range azimuthal correlations that are attributed to the hydrodynamical evolution of the equilibrated medium~\cite{Ollitrault:1992bk}. This interpretation has been widely tested in heavy-ion collisions~\cite{PHENIX:2004vcz, PHOBOS:2004zne, STAR:2005gfr, ALICE:2010suc}. A similar collective phenomenon has been observed in medium-sized ion collisions, like Oxygen-Oxygen~\cite{ALICE:2025luc, CMS:2025tga} and small systems in high-multiplicity events, like proton-lead~\cite{ATLAS:2014qaj} and proton-proton~\cite{CMS:2016fnw} collisions. While another sign of QGP formation, jet quenching, has been observed in the former~\cite{CMS:2025bta}, it has not been observed in the latter~\cite{CMS:2015ved}, raising the puzzle of whether a deconfined system of quarks and gluons is actually formed in such small colliding systems.

Even though hydrodynamical simulations can reproduce the correlations observed in experiments, the applicability of relativistic hydrodynamics has been questioned for small systems. If the origin of the azimuthal correlations in small systems is not hydrodynamical, then other mechanisms should arise to produce them. It has been proposed that these correlations can be formed at the time of the collision, as described within the Color Glass Condensate (CGC) framework (see~\cite{Altinoluk:2020wpf} for a recent review), or due to interference effects between valence quarks~\cite{Li:2023vdj}. The presence of this effect can be disregarded in heavy-ion collisions, since the final state interaction during the thermalization will wash out the initial anisotropies. However, as the system size decreases, the thermalization time becomes comparable to the lifetime of the QGP, and the pre-equilibrium stages are expected to be eventually reachable in experimental observables. 

In the present work, we study how these initial anisotropies relax over time due to the final-state interactions present in the BEDA framework. The structure of the paper is as follows. In Section~\ref{sec:BEDA}, we review the Boltzmann Equation in Diffusion Approximation as our tool to study thermalization and isotropization. In Section~\ref{sec:Azim1} we study how azimuthal anisotropies relax in terms of fully integrated harmonic (Fourier) coefficients in order to extract some general features of the process. In Section~\ref{sec:Azim2}, we also study such isotropization but as a function of the transverse momenta, $p_T$, since initial state calculations predict initial anisotropies as a function of $p_T$ to take some non-trivial shape. We also tentatively perform some phenomenological studies to understand the implications of the final state interactions on the observed $v_n$ coefficients in small colliding systems. Finally, in Section~\ref{sec:Conclusions}, we summarize and conclude our results. Two appendices complement the current work. In Appendix~\ref{sec:app_others}, we add to the qualitative evolution explained in Section~\ref{sec:Azim1} for $2\leftrightarrow2$ interactions a similar analysis for the case of the $1\leftrightarrow2$ collision kernel. Finally, since this work requires the numerical simulation of the BEDA, we have developed a GPU-based numerical algorithm that is explained in detail in Appendix~\ref{sec:app_numerics}. 

\section{The Boltzmann Equation in Diffusion Approximation}\label{sec:BEDA}

Under the assumption of  homogeneity in the transverse plane and longitudinal boost-invariant expansion, the QCD Boltzmann Equation at leading order can be written as~\cite{Baier:2000sb, Arnold:2002zm}
\begin{eqnarray}\label{eq:BE}
    \left(\partial_\tau - \frac{p_z}{\tau} \partial_{p_z}\right) f^a(\tau;\mathbf{p}) = C^a_{1\leftrightarrow2}[f] + C^a_{2\leftrightarrow2}[f]~,
\end{eqnarray}
where the $a$ index indicates that we are considering quarks and gluons, and $\mathbf{p}$ is the three-momentum vector. The complete leading-order treatment requires computing the collision integrals as in the Effective Kinetic Theory~\cite{Arnold:2002zm}, and it is possible to obtain a simplified version after taking the diffusion approximation in the $2\leftrightarrow2$ kernel and restricting the inelastic interactions to the deep LPM regime. This is the case of the Boltzmann Equation in Diffusion Approximation (BEDA)~\cite{Cabodevila:2023htm, BarreraCabodevila:2025vir}. 

Let us briefly review how these collision kernels can be written under these approximations. In our previous work~\cite{Cabodevila:2023htm}, we have written the collision kernels for a general case with asymmetry between quarks and antiquarks. However, we will restrict ourselves here to systems with no quarks at the initial time and, therefore, no asymmetry between fermions and antifermions will arise. From now on, $f\equiv f_g$ and $F \equiv f_q$ refer to the distribution functions of gluons and quarks, respectively.

On one hand, under the diffusion approximation, that is, restricting the $2\leftrightarrow2$ scatterings to processes with small momentum transfer, the corresponding collision integral reduces to a Fokker-Planck equation plus an additional source term~\cite{Mueller:1999pi, Baier:2000sb, Hong:2010at, Blaizot:2013lga, Blaizot:2014jna},
\begin{align}
C^a_{2\leftrightarrow2}=\frac{1}{4}\hat{q}_a\nabla_{{\bm p}} \cdot\left[  \nabla_{{\bm p}}f^a  + \frac{{\bm v}}{T^*}f^a(1+\epsilon_a f^a)\right]+\mathcal{S}_a,
\label{eq:C2to2}
\end{align}
with $\epsilon_g=+1$ and $\epsilon_q=-1$ encoding the quantum statistics and
\begin{align}\label{eq:Sqg}
   \mathcal{S}_{q} =\frac{2\pi\alpha_s^2  C_F^2 \mathcal{L}}{ p}\mathcal{I}_c\bigg[ f (1 - F ) - F ( 1 + f ) \bigg],\qquad
   \mathcal{S}_g=-\frac{N_f}{C_F}\mathcal{S}_{q}.
\end{align}
Here, $\alpha_s$ is the coupling constant, $C_F=4/3$ is the invariant Casimir in the fundamental representation of color $SU(N_c)$, and $N_f$ is the number of active quarks in the model. The values $\hat{q}_a$, $T_*$ and $\mathcal{I}_c$ correspond to some integral moments of the distribution functions of gluons and quarks. The first one is the jet quenching parameter (broadening coefficient), $\hat{q} = C_a \hat{\bar{q}}$~\cite{Baier:1996sk}, with
\begin{align}
    \hat{\bar{q}} \equiv 8\pi \alpha_s^2\mathcal{L}\int\frac{d^3\bm p}{(2\pi)^3}
    \left[N_c f (1+f) + N_f F (1 - F) \right]~.
\end{align}
The second one is the effective temperature, which can be shown to match the temperature once the system has equilibrated
\begin{align}\label{eq:Ts}
    T_* \equiv \frac{\hat{q}_A}{\alpha_s N_c \mathcal{L} m_D^2}~.
\end{align}
It relates the value of the quenching parameter and the squared screening mass, defined as
\begin{align}
    m_D^2 \equiv 16\pi\alpha_s\int\frac{d^3 \bm{p}}{(2\pi)^3}\frac{1}{p} (N_c f + N_f F).
\end{align}  
Finally, $\mathcal{I}_c$ can be interpreted as a coefficient that determines the ratio of the $g\leftrightarrow q$ conversion due to $2\leftrightarrow2$ interactions. It is defined as
\begin{align}\label{eq:IcIcb}
    \mathcal{I}_c = \int \frac{d^3\bm{p}}{(2\pi)^3} \frac{1}{p} (f + F).
\end{align}
Some of these expressions also involve a logarithm that arises from the integration of the IR divergences~\cite{Mueller:1999fp}. In our case, following the discussion of Ref.~\cite{BarreraCabodevila:2025vir}, we set it to vary in time according to the expression
\begin{align}
\label{eq:Lchoice}
    \mathcal{L} = \ln \bigg(\frac{\hat{\bar{q}} t_{\text{br}}(\bar{p})}{m_D^2}\bigg) = \ln \bigg(\frac{\sqrt{\hat{\bar{q}}\bar{p}}}{\alpha_s m_D^2}\bigg)~.
\end{align}  

On the other hand, the $1\leftrightarrow2$ collision kernel is computed in the deep LPM regime for collinear splittings:
\begin{align}
    C^a_{1\leftrightarrow2} = &\int_0^1 dx\sum\limits_{b,c}\bigg[\frac{1}{x^3}\frac{\nu_c}{\nu_a}C^c_{ab}({\bm p}/{x};{\bm p},{\bm p}(1-x)/x)-\frac{1}{2}C^a_{bc}(\bm p;x{\bm p},(1-x){\bm p})\bigg]~.
    \label{eq:inel_kern}
\end{align}
In this equation, the integration is carried out over the energy fraction $x$, and the sum involves all possible $a\leftrightarrow bc$ processes allowed by the QCD interaction vertices. The information of the $a \to bc$ splitting is encoded in
\begin{align}
    C^{a}_{bc}(\bm p;x{\bm p},(1-x){\bm p})\equiv\frac{dI_{a\to bc}(p)}{dxdt}\mathcal{F}^a_{bc}(\bm p;x{\bm p},(1-x){\bm p})~,
    \label{eq:inel_kern2}
\end{align}
with $dI_{a\to bc}/dx dt$ the splitting rate in the deep LPM regime\footnote{The detailed expressions as a function of $\hat{q}_a$ are summarized in Ref.~\cite{Cabodevila:2023htm}.}~\cite{Baier:1998kq, Arnold:2008zu}; and the statistical factor
\begin{align}
    \mathcal{F}^a_{bc}(\bm p;{\bm l},{\bm k})\equiv f^a_{\bm p}(1+\epsilon_b f^b_{\bm l})(1+\epsilon_c f^c_{\bm k})-f^b_{\bm l}f^c_{\bm k}(1+\epsilon_a f^a_{\bm p})~.
    \label{eq:inel_kern3}
\end{align}

\section{Relaxation of azimuthal anisotropies}\label{sec:Azim1}

Color Glass Condensate calculations have shown that it is possible that the system created after the heavy ion collision can be produced with initial azimuthal anisotropies. Previous studies of thermalization have disregarded the effect that these initial anisotropies might have on equilibration~\cite{Baier:2000sb, Kurkela:2015qoa, BarreraCabodevila:2025vir} since they are small and are expected to be quickly washed out due to final state interactions. Since they are not expected to play any role in the thermalized medium, the anisotropies measured in the experiments are exclusively explained by the collective dynamics of the fluid resulting from the collision. However, in smaller colliding systems, the system might not have enough time to relax these initial anisotropies, so they can have their imprint in the measured azimuthal correlations. 

In this section and in the following, we study in detail how these initial anisotropies evolve in time due to the final state interactions included in the Boltzmann Equation in Diffusion Approximation. As mentioned in Section~\ref{sec:BEDA}, we assume longitudinal boost-invariance and homogeneity in the transverse plane so that we can neglect spatial gradients.

\subsection{Initial conditions and Fourier coefficient evolution}

In this work, we assume that the initial condition is given by the CGC-inspired initial condition~\cite{Kurkela:2015qoa, Kurkela:2018oqw} supplemented by an azimuthal anisotropy as a Fourier series,
\begin{eqnarray}\label{eq:InitCond}
    f_0(\mathbf{p}) = f_{0}(p, \cos\theta) \left[ 1 + 2 \sum_n v_n(p, \cos \theta) \cos(n\phi) \right]~.
\end{eqnarray}
Here, the distribution $f_0(p,\cos\theta)$ is designed to capture some of the relevant scales of the system at the time when the kinetic evolution starts to be applicable:
\begin{eqnarray}
    f_0 = \frac{A}{4\pi N_c \alpha_s}\frac{e^{-\frac{2}{3}\left( (p_z \xi)^2 + p_T^2 \right) / Q_0^2}}{\sqrt{\left( (p_z \xi)^2 + p_T^2 \right) / Q_0^2}}~.
\end{eqnarray}
In this expression, $\xi$ is a parameter used to quantify the initial anisotropy of the distribution function, and $Q_0$ is the typical transverse momentum of the partons at this time. For heavy-ion collisions at LHC energies, it has been estimated~\cite{Lappi:2011ju} that $Q_0 \approx 1.8 Q_s$, with $Q_s$ the saturation scale. In this work, we take $\tau_0 = Q_s^{-1}$. The constant $A$ is used to match some phenomenological energy density at the initial time~\cite{Kurkela:2015qoa, Kurkela:2021ctp}.

In general, the $v_n$ coefficients may depend on $p$ and $\cos \theta$, but in this section we restrict ourselves to the simplest scenario and treat them as constants. In the next section, we will assume some $p_T$ dependence relevant for phenomenology studies. Also, we do not consider any initial population of quarks throughout this work, even though they can be produced with the processes explained in Section~\ref{sec:BEDA}; see relevant discussions in~\cite{Cabodevila:2023htm, BarreraCabodevila:2025vir}.

To investigate the isotropization of the azimuthal anisotropies, we recall the relation between the distribution function and the produced particle spectra,
\begin{equation}
    \frac{1}{\tau}\frac{d^3N}{d^2\mathbf{p}_T dy d\eta}\Bigg |_\tau \propto\int d^2\mathbf{x_\perp} f(\tau; \eta; \mathbf{p}_T, p_T\sinh (y-\eta))~.
\end{equation}
As we focus on early-time dynamics, the system is assumed to be homogeneous in the transverse plane. In this case, we use $\tau$ to mimic the effect of a finite transverse size that, at time $\tau$, all partons in the system will propagate freely towards the detector, so that we can match the value of our distribution function with the particle spectra. Under this assumption, the $\mathbf{x}_\perp$ integral becomes an overall factor. Now, since we are mostly interested in the $p_T$ dependence of the azimuthal anisotropies integrated over a finite range in $\eta$, as is typically done in experimental analyses, we integrate over the rapidities,
\begin{align}
\begin{split}
    \frac{1}{\tau p_T}\frac{d^3N}{dp_T d\phi}\Bigg |_\tau &\propto  \int_{\eta_{min}}^{\eta_{max}}d\eta\int dy  \int d^2\mathbf{x_\perp} f(\tau; \eta; \mathbf{p}_T, p_T\sinh (y-\eta)) = \\
    &= \int_{\eta_{min}}^{\eta_{max}}d\eta \int dy'  \int d^2\mathbf{x_\perp} f(\tau; \eta; \mathbf{p}_T, p_T\sinh (y'))~.
\end{split}
\end{align}
In the second line, we used the fact that the momentum rapidity integration is performed from $-\infty$ up to $\infty$ and performed the change of coordinates $y\to y'+\eta$. Now, using $p_z=p_T \sinh (y)$, the $y$ integral transforms into an integral over $p_z$. By dropping overall factors that will cancel in later expressions, we obtain
\begin{eqnarray}\label{eq:pespectra}
    \frac{1}{\tau p_T}\frac{d^3N}{dp_T d\phi}\Bigg |_\tau \propto \int dp_z f(\tau; \mathbf{p}_T, p_z)~.
\end{eqnarray}

The time evolution of collective modes with azimuthal angle dependence can be described in terms of Fourier coefficients. We follow the evolution of these Fourier coefficients in the event plane (EP), defined from the particle spectra obtained in Eq.~\eqref{eq:pespectra}, as
\begin{eqnarray}\label{eq:vn_differential}
    v_n^a(\tau, p_T) = \frac{ \int \frac{d\phi}{(2\pi)^3} \frac{dN^a}{dp_T d\phi}\Big |_\tau \cos \left[ n (\phi - \psi_n^a) \right]}{\int \frac{d\phi}{(2\pi)^3} \frac{dN^a}{dp_T d\phi}\Big |_\tau}~,
\end{eqnarray}
with the EP angle,
\begin{eqnarray}
    \psi_n^a (\tau)\equiv \frac{1}{n} \arctan \frac{\int \frac{d^3\mathbf{p}}{(2\pi)^3} \frac{dN^a}{dp_T d\phi}\Big |_\tau \sin (n \phi)}{\int \frac{d^3\mathbf{p}}{(2\pi)^3} \frac{dN^a}{dp_T d\phi}\Big |_\tau \cos (n \phi)}~.
\end{eqnarray}
Because the ansatz for the initial condition in Eq.~\eqref{eq:InitCond} does not include different event-plane orientations, we always take $\psi_n = 0$ in our results. Nevertheless, we keep it explicit for completeness.

As commented previously, because of the ratio structure, the overall factors associated with the integration over the transverse area and rapidity window cancel. Therefore, we can directly replace the particle spectrum in these expressions by $\int dp_z\, f^a(\tau;\mathbf{p})$, and we will do so in the following. The superscript $a$ denotes the parton species considered, namely quarks and gluons. We can also define the distribution for the full system by noting that
\begin{equation}
f^{\mathrm{all}} \equiv \nu_g f^g + 2N_f\nu_qf^q~,
\end{equation}
with degeneracy factors $\nu_g = 16$ and $\nu_q = 6$ for a system with quark–antiquark symmetry.

Eq.~\eqref{eq:vn_differential} contains detailed dynamical information on how the anisotropy evolves in phase space. For this reason, in this section we focus on the momentum-integrated Fourier coefficients,
\begin{eqnarray}\label{eq:vn_integrated}
    v_n^a \equiv \frac{\int \frac{d^3\mathbf{p}}{(2\pi)^3} f^a(\mathbf{p}) \cos \left[ n (\phi - \psi_n) \right]}{\int \frac{d^3\mathbf{p}}{(2\pi)^3} f^a(\mathbf{p})}~.
\end{eqnarray}
Given the previous definitions, the variation of the $v_n$ coefficient in Eq.~\eqref{eq:vn_differential} induced by each collision kernel $\mathcal{C}$ can be written as\footnote{From now on we do not distinguish between particle species; we include both quarks and antiquarks.}
\begin{align}\label{eq:deriv_coef}
\begin{split}
    (\partial_\tau v_n)_{\mathcal{C}} \equiv \dot{v}_n(p_T) = \frac{1}{N^2} \int \frac{d\phi dp_z}{(2\pi)^3} \left[  cos(n(\phi-\psi_n)) \left( \dot{f}N - f \dot{N} \right) \right. ~~\\ \left. -N \sin (n(\phi-\psi_n))\dot{\psi}_n f \right]~,
\end{split}
\end{align}
where
\begin{eqnarray}\label{eq:deriv_coef_2}
    N \equiv \int\frac{d\phi dp_z}{(2\pi)^3} f ~,\quad \dot{N} = \int \frac{d\phi dp_z}{(2\pi)^3} \dot f = \int \frac{d\phi dp_z}{(2\pi)^3}  \mathcal{C}~,
\end{eqnarray}
and
\begin{eqnarray}
    \dot \psi_n = \frac{1}{n} \frac{\int_{\mathbf{p}} \sin(n\phi)\mathcal{C} \int_{\mathbf{p}} \cos(n\phi)f - \int_{\mathbf{p}} \sin(n\phi)f \int_{\mathbf{p}} \cos(n\phi)\mathcal{C} }{\left[\int_{\mathbf{p}} \sin(n\phi)f\right]^2 + \left[\int_{\mathbf{p}} \cos(n\phi)f\right]^2}~.
\end{eqnarray}
The corresponding variation for the integrated coefficients in Eq.~\eqref{eq:vn_integrated} can be easily obtained by replacing $\int \frac{dp_z\, d\phi}{(2\pi)^3}$ with $\int \frac{d^3\mathbf{p}}{(2\pi)^3}$ in the expressions above.

For the following study, we initialize the system with a constant value $v_n(\tau=\tau_0)=0.25$ for a single harmonic $n=2,3,4$. We consider two types of evolution: one with $N_f=0$ and another with $N_f=3$ for each case\footnote{In both cases, the system is initially populated by gluons, while in the $N_f=3$ case, fermions can be produced dynamically during the thermalization process.}. All data shown in this work are generated using the algorithm described in Appendix~\ref{sec:app_numerics}, with $N_p=N_{v_z}=N_\phi=64$. The momentum grid is logarithmic, with $p_{\min}=0.02\,Q_s$ and $p_{\max}=10\,Q_s$, while the $v_z$ and $\phi$ grids are linear. The coupling constant is chosen as $\lambda=4\pi N_c \alpha_s = 10$, consistent with heavy-ion phenomenology.

\subsection{Qualitative evolution of momentum isotropization}

Some qualitative insights into azimuthal isotropization can be obtained from a simple inspection of the collision kernels and their effect on the time evolution of the fully integrated $v_n$ coefficients. In this case, two main processes can be identified: the relaxation of the initial $v_n$ coefficients and the generation of higher-order $v_n$ coefficients. We now discuss both qualitatively.

Since the Fokker--Planck term of the $2\leftrightarrow2$ collision kernel conserves particle number density, the integrated version of Eq.~\eqref{eq:deriv_coef} simplifies considerably. As a first example, we compute the variation of the $v_n$ coefficient induced by $C_{2\leftrightarrow 2}\big|_{\mathrm{FP}}$ by inserting the ansatz
\begin{eqnarray}\label{eq:ansatz}
    f^a(\mathbf{p};\tau) = \tilde{f}(p,\cos \theta; \tau) \left[ 1 + 2\sum_n v_n(p,\cos\theta; \tau) \cos (n\phi) \right]
\end{eqnarray}
in Eq.~\eqref{eq:deriv_coef} and obtain\footnote{In this calculation, a term of the form
\begin{equation*}
    \sum_{k,l} v_k v_l \int \frac{d\phi}{2\pi}\,
    \cos(n\phi)\cos(k\phi)\cos(l\phi)
\end{equation*}
appears. Since $n,k,l>0$, as they label harmonic modes, this integral can be evaluated as
\begin{equation*}
    \int\frac{d\phi}{2\pi}\cos(n\phi)\cos(k\phi)\cos(l\phi)
    =\frac{1}{4}(\delta_{n,k+l}+\delta_{k,l+n}+\delta_{l,n+k})
\end{equation*} and that is where the term in the second line comes from.}
\begin{align}
\label{eq:dtvnel}
    \partial_\tau v_n(\tau) &= \frac{\hat{q}}{4n_g} \int\frac{p^2dp~d\cos \theta}{(2\pi)^2} \bigg\{\bigg[-\frac{n^2\tilde{f}}{2p^2(1-\cos^2\theta)}
    +\frac{1}{T_*} \bigg(\frac{2}{p}+\partial_p\bigg)(\tilde{f}+\tilde{f}^2)\bigg]v_n^2(p,\cos\theta)\notag\\
    &+\frac{1}{T_*} \bigg( \frac{2 \tilde{f}^2}{p} + \partial_p \tilde{f}^2 \bigg) \sum_{k,l}\frac{v_k(p,\cos\theta) v_l(p,\cos\theta)}{4}(\delta_{n,k+l}+\delta_{k,l+n}+\delta_{l,n+k})\bigg\}.
\end{align}
Here we have used the fact that $(\partial_t \psi_n)_{2\leftrightarrow2} = 0$. This result is derived for a single species, but it can be straightforwardly extended to obtain the overall behavior due to $2\leftrightarrow2$ interactions because contributions from the source term cancel between quarks and gluons.

From Eq.~\eqref{eq:dtvnel}, we identify two distinct effects in the evolution of the $v_n$ coefficients during isotropization. On one hand, the first line gives rise to a relaxation-like contribution\footnote{Although the $v_n$ appearing here are not the fully integrated coefficients, one can see from their definition in Eq.~\eqref{eq:vn_integrated} that they are related to the coefficients appearing in the ansatz of Eq.~\eqref{eq:ansatz} as
\begin{eqnarray*}
    v_k = \,
    \frac{1}
    {n}\int dp\, p^2 \cos\theta\, \tilde{f}(p,\cos\theta)\, v_k(p,\cos\theta)~,
\end{eqnarray*}
where $n$ is the number density. Thus, a nonzero integrated $v_k$ requires $v_k(p,\cos\theta)\neq 0$.} The rapid expansion, combined with the energy flux toward the infrared induced by $1\leftrightarrow2$ processes, reduces the occupancy of the hard sector, so the first term in the brackets governs the relaxation of the coefficients. This term defines a clear hierarchy in the isotropization dynamics: higher-order harmonics (larger $n$) relax faster than lower-order ones. On the other hand, the last line shows that higher-order coefficients can be generated dynamically during isotropization. In particular, starting from an initial condition with a nonzero $v_n$, all harmonics of the form $v_{in}$ (with $i=1,2,3,\dots$) can be generated through this contribution.

Even though the radiation is collinear, the $1\leftrightarrow2$ collision integral also affects the evolution because particle flow in phase space modifies the integrals entering Eq.~\eqref{eq:vn_integrated}. To complement this analysis, we present the contributions of inelastic interactions and the source term of $C_{2\leftrightarrow2}$ in Appendix~\ref{sec:app_others}. From the statistical term in the expression of $C_{1\leftrightarrow2}$, one observes $f^2$ terms that generate contributions to $(\partial_t v_n)_{1\leftrightarrow2}$ proportional to an integral with three cosine factors, similar to those appearing in Eq.~\eqref{eq:dtvnel}. Thus, inelastic processes also modify the evolution of the fully integrated harmonic coefficients. Ultimately, both $1\leftrightarrow2$ and $2\leftrightarrow2$ collision integrals render the time evolution of the $v_n$ coefficients a highly nontrivial process that needs to be studied numerically for a complete understanding.

\subsection{Numerical results}

\begin{figure}
    \centering
    \includegraphics[width=0.99\linewidth]{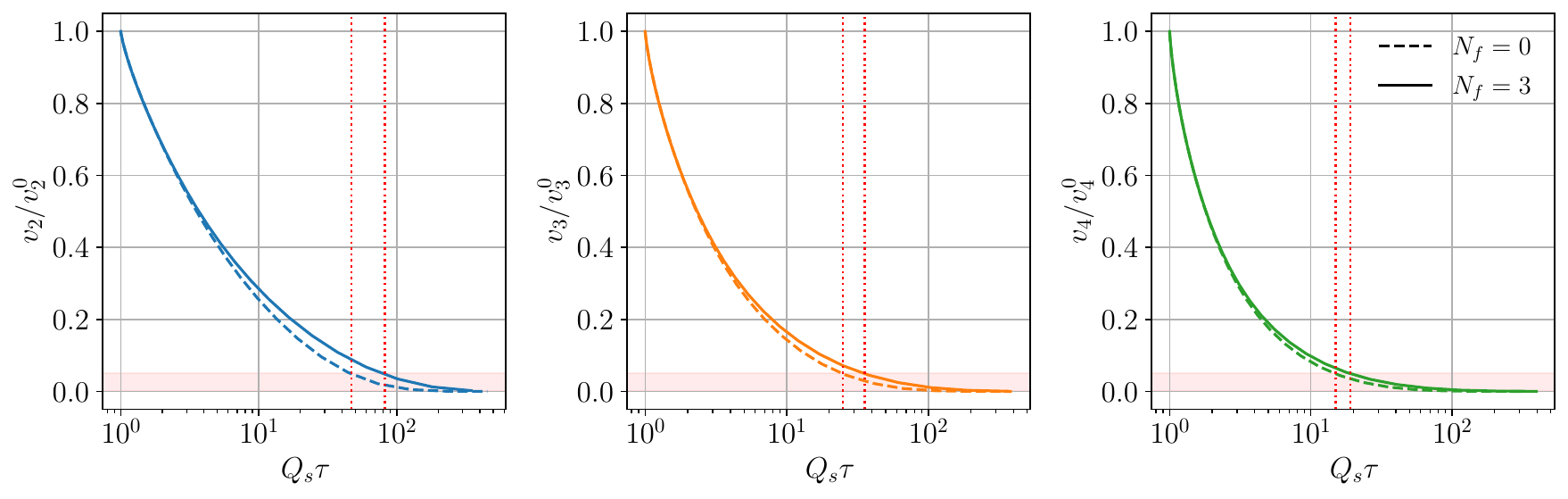}
    \caption{Time evolution of the $v_n$ coefficients defined in Eq.~\eqref{eq:vn_integrated} for initially nonzero $n=2,3,4$ (from left to right). The ratios of $v_n$ to their corresponding values at $\tau=\tau_0$, denoted by $v_n^0$, are shown. Results are presented for both $N_f=0$ and $N_f=3$. The red dashed vertical lines indicate the time at which the isotropization condition of Eq.~\eqref{eq:isocond} is reached.
}
    \label{fig:vns_integrated_full}
\end{figure}

The previous claims are verified by numerical solutions of Eq.~\eqref{eq:BE}. In Fig.~\ref{fig:vns_integrated_full}, we show the time evolution of the $v_n$ coefficients for each initial anisotropy. As expected from the analysis in Fig.~\eqref{eq:dtvnel}, a clear hierarchy in the relaxation time as a function of $n$ emerges. In addition, we observe a systematic delay in the isotropization for systems with active quarks ($N_f=3$).

We define the azimuthal isotropization time as the time at which the leading harmonic coefficient decreases to $5\%$ of its initial value\footnote{Due to the hierarchy in relaxation times of the harmonic coefficients, the leading contribution comes from the lowest nonzero $n$, which isotropizes most slowly.},
\begin{eqnarray}\label{eq:isocond}
    \frac{v_n(\tau_{\mathrm{iso}})}{v_n(\tau_0)} = 0.05~.
\end{eqnarray}
In our simulations, we find that for the $N_f=0$ scenario, $\tau_{\mathrm{iso}} \approx 47\,Q_s^{-1}$, $25\,Q_s^{-1}$, and $15\,Q_s^{-1}$ for initial $v_{2,3,4}\neq 0$, respectively. For $N_f=3$, we obtain $\tau_{\mathrm{iso}} \approx 82\,Q_s^{-1}$, $36\,Q_s^{-1}$, and $19\,Q_s^{-1}$, in the same order. Thus, the delay in isotropization between the pure-gluon system and the quark–gluon plasma decreases as $n$ increases. This is a natural result, since fermionic degrees of freedom are less relevant at early times. For larger $n$, isotropization occurs more rapidly and predominantly within the gluonic sector, while quarks are produced with a much smaller anisotropy. Overall, higher-$n$ harmonics isotropize faster, and the role of quarks in the isotropization dynamics becomes less significant. 

\begin{figure}
    \centering
    \includegraphics[width=0.4\linewidth]{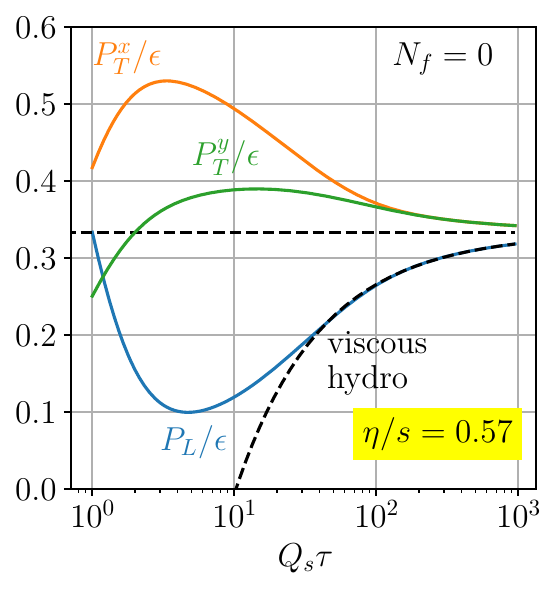}
    \includegraphics[width=0.4\linewidth]{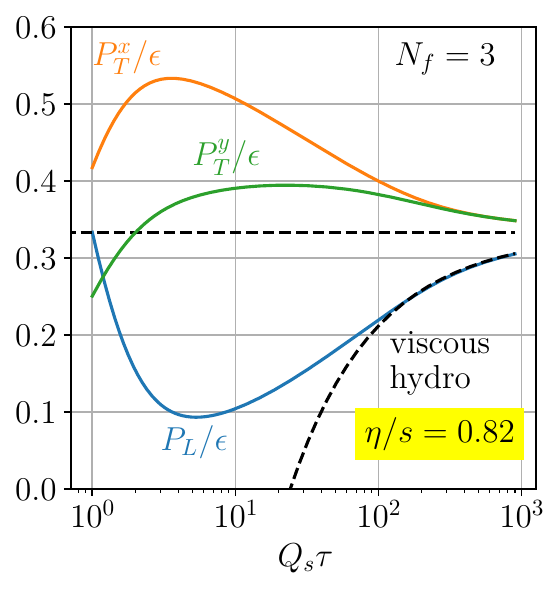}
    \caption{Ratios of the pressures in each direction with respect to the energy density for the initial condition with $v_2=0.25$. The left panel corresponds to $N_f=0$ and the right panel to $N_f=3$.}
    \label{fig:pressures}
\end{figure}

The isotropization time of the initial azimuthal anisotropies is very close to the hydrodynamization time. In Figure~\ref{fig:pressures} we show the pressures ratios with respect to the energy density for the three directions, with the $P_L/\epsilon$ fitting the hydrodynamic attractor at late times. In this case, the hydrodynamization time is at $\tau_{\mathrm{hydro}} \approx 28Q_s^{-1}$ for $N_f=0$ and $\tau_{\mathrm{hydro}} \approx 84Q_s^{-1}$ for $N_f=3$. In both cases, the hydrodynamization times are comparable to $\tau_{\mathrm{iso}}$, so the effects of the initial azimuthal anisotropies may survive until the onset of the hydrodynamical regime.

\begin{figure}
    \centering
    \includegraphics[height=0.4\linewidth]{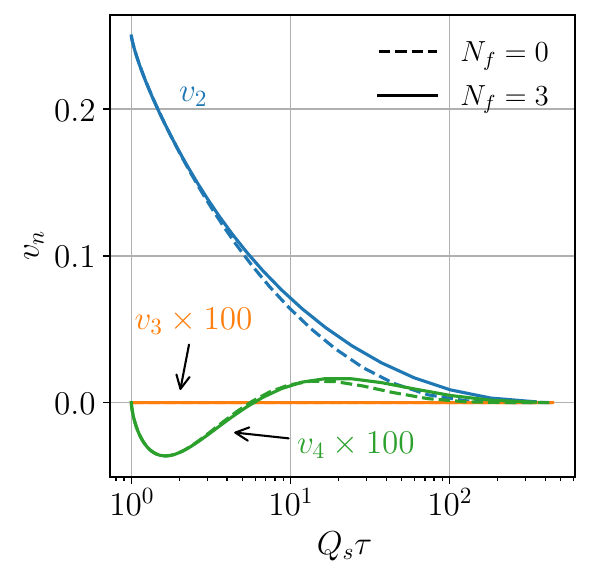}
    \includegraphics[height=0.4\linewidth]{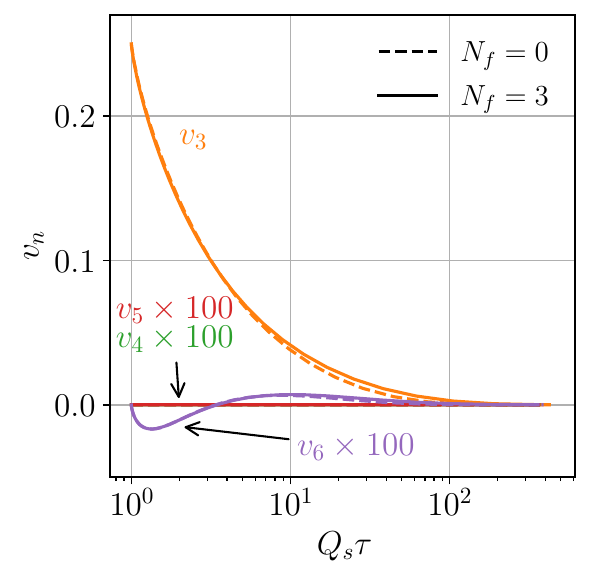}
    \caption{Time evolution of different $v_n$ coefficients for two initial conditions: initial $v_2\neq0$ (left) and initial $v_3\neq0$ (right). Results for both $N_f=0$ and $N_f=3$ are displayed. The higher-order harmonic coefficients have been rescaled to make their evolution visible.}
    \label{fig:vnprod_full}
\end{figure}

The generation of higher-order harmonics can be observed in both panels of Fig.~\ref{fig:vnprod_full}. As expected, for an initial condition with $v_2\neq0$, $v_3$ remains zero, while a non-zero $v_4$ is generated during the isotropization process. The evolution of this coefficient is non-trivial. It first decreases toward negative values and reaches a minimum at $Q_s\tau\approx2$. It then increases, reaching a maximum at $Q_s\tau\approx20$, before finally relaxing toward zero as isotropization is achieved.

A similar behavior is observed for the case with $v_3\neq0$. In this case, we see that both $v_4$ and $v_5$ remain zero throughout the isotropization process, while a non-zero $v_6$ is generated. Initially, it becomes negative and reaches a minimum at $Q_s\tau\approx1.5$, whereas its maximum occurs at $Q_s\tau\approx10$. The faster evolution of the higher-order harmonic compared to the previous case is related to the hierarchy of relaxation times discussed in the previous section. That is, the evolution is faster at early times because the relaxation of the leading harmonic is also faster. Once the higher-order coefficient becomes non-zero, the relaxation term given by the first line of Eq.~\eqref{eq:dtvnel} drives it toward isotropization more rapidly than the lower-order harmonics.

\begin{figure}
    \centering
    \includegraphics[width=0.99\linewidth]{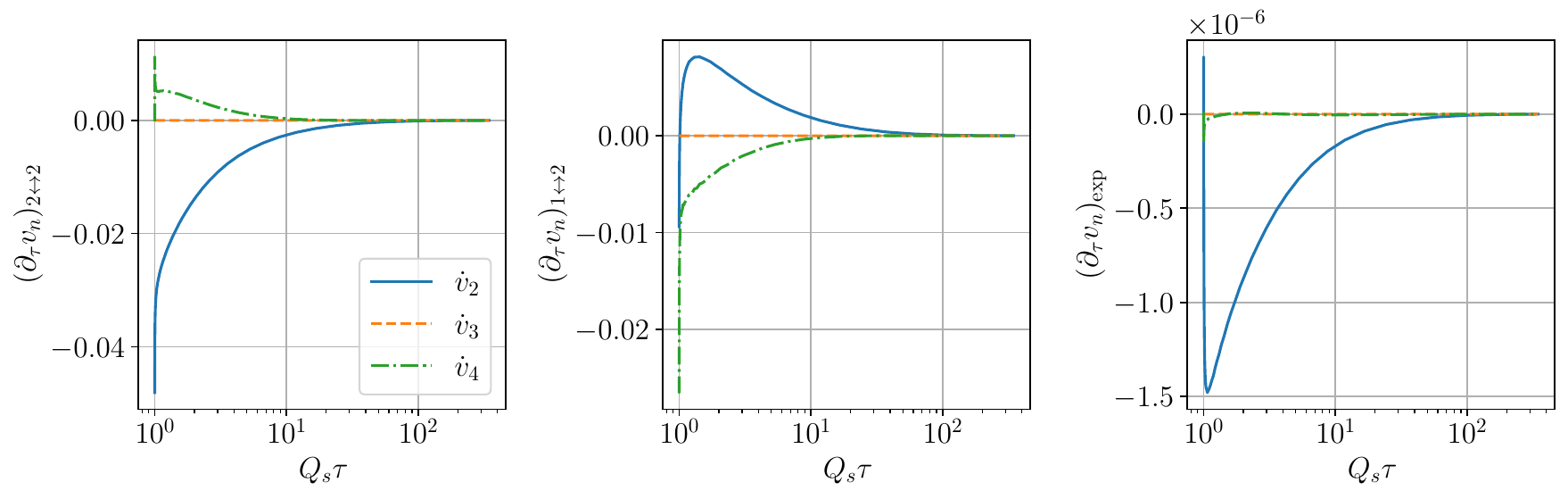}
    \caption{Variation of $v_2$, $v_3$, and $v_4$ for a system with the initial condition $v_2\neq0$, due to the $2\leftrightarrow2$ collision kernel (left), the $1\leftrightarrow2$ collision kernel (center), and the expansion term (right), for the $N_f=3$ scenario.}
    \label{fig:vnsderivs}
\end{figure}

Let us focus on the case with the initial condition $v_2\neq0$ to explore the contribution of each collision kernel to isotropization. These contributions can be computed using Eq.~\eqref{eq:deriv_coef} and are shown for this case in Fig.~\ref{fig:vnsderivs}. The contribution from the expansion to the evolution of the harmonic coefficients is clearly much smaller than that of the collision kernels and can therefore be neglected in the evolution. Thus, let us discuss in detail how the $1\leftrightarrow2$ and $2\leftrightarrow2$ processes contribute to azimuthal isotropization.

First, the $2\leftrightarrow2$ kernel is the dominant contribution to the relaxation of the leading harmonic at early times. The $1\leftrightarrow2$ processes quickly oppose the elastic interactions, tending to increase the value of $v_n$. Numerically, one can verify that the sum of elastic and inelastic contributions is always negative, which explains why we always observe the same relaxation trend in Fig.~\ref{fig:vnprod_full}. The tendency of the inelastic processes to increase $v_n$ is related to our definition of the harmonic coefficients. The $1\leftrightarrow2$ processes are collinear, and one should not expect a rearrangement of the angular structure of the distribution function due to them. However, in the computation of the $v_n$ coefficients, we integrate over the modulus of the momentum to match the particle spectra. Since $1\leftrightarrow2$ processes redistribute energy and number density over the full momentum range at a fixed solid angle, this leads to a change in the extracted $v_n$.

Second, none of the kernels significantly affects the $v_3$ coefficient, as expected, but they do generate a non-zero $v_4$. This has already been discussed in the previous section, but we now quantify the effect. In contrast with the leading-harmonic evolution, the contribution of the $1\leftrightarrow2$ processes is dominant over the $2\leftrightarrow2$ kernel at early times, generating a negative $v_4$. As the system evolves, both contributions become comparable, and eventually the elastic contribution overcomes the inelastic one, driving $v_4$ toward positive values. Finally, the coefficient relaxes mainly due to the $2\leftrightarrow2$ interactions\footnote{The final relaxation is always driven by the $2\leftrightarrow2$ collision kernel, since it is the one that governs the angular structure of the distribution function. Once the (faster) $1\leftrightarrow2$ splittings and mergings have built up a thermal-like distribution at each solid angle, the elastic interactions isotropize the remaining distribution.}.

\section{Momentum dependent azimuthal anisotropies}\label{sec:Azim2}

In the previous section, we did not explore the momentum dependence of the $v_n$ coefficients in order to develop a more general understanding of how the initial azimuthal anisotropies relax. However, initial-state calculations have shown that these anisotropies exhibit a clear dependence on the transverse momentum~\cite{Altinoluk:2020wpf}. These studies typically find a peaked structure around $p_T \sim Q_s$ for the momentum dependence of $v_n(p_T)$. With this in mind, we model the harmonic coefficients in the initial condition~\eqref{eq:InitCond} as
\begin{eqnarray}\label{eq:init_vnpt}
    v_n(p_T) = \tilde{v}_n\frac{p_T}{Q_s} e^{-|p_T|/Q_s}~.
\end{eqnarray}
Now, let us consider this dependence for the $v_n$ coefficients at the initial time and explore its time evolution. Since we aim to understand how the shape of the anisotropy evolves with time, we start with the simplest case, taking $\tilde{v}_2 = 0.25$ and $\tilde{v}_n = 0$ for $n \neq 2$. First, we examine the effect of final-state interactions on the $p_T$-dependent coefficients, defined as
\begin{eqnarray}\label{eq:vn_ptdifferential}
    v_n^a(p_T,\tau) = \frac{\int \frac{dp_zd\phi}{(2\pi)^3} f^a(\mathbf{p}) \cos \left[ n (\phi - \psi_n^a) \right]}{\int \frac{dp_zd\phi}{(2\pi)^3} f^a(\mathbf{p})}~.
\end{eqnarray}
Then, we will tune the initial conditions to see how these interactions allow us to mimic experimental results.

\subsection{Evolution of the harmonic coefficients}

\begin{figure}
    \centering
    \includegraphics[trim=0 0 3.2cm 0, clip, height=0.26\linewidth]{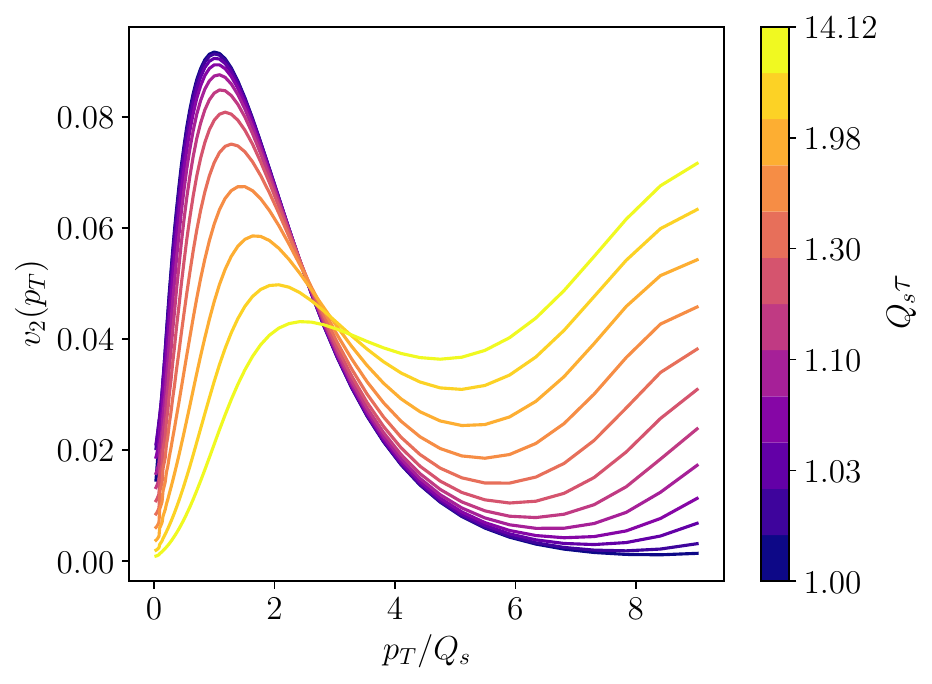}
    \includegraphics[trim=0 0 3.2cm 0, clip, height=0.26\linewidth]{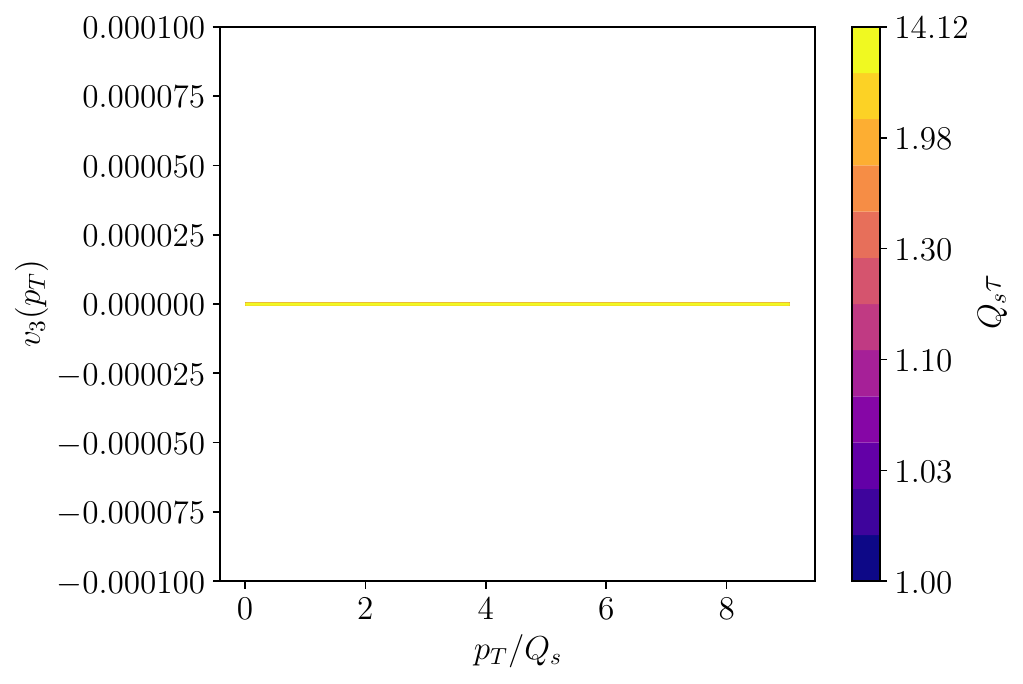}
    \includegraphics[height=0.26\linewidth]{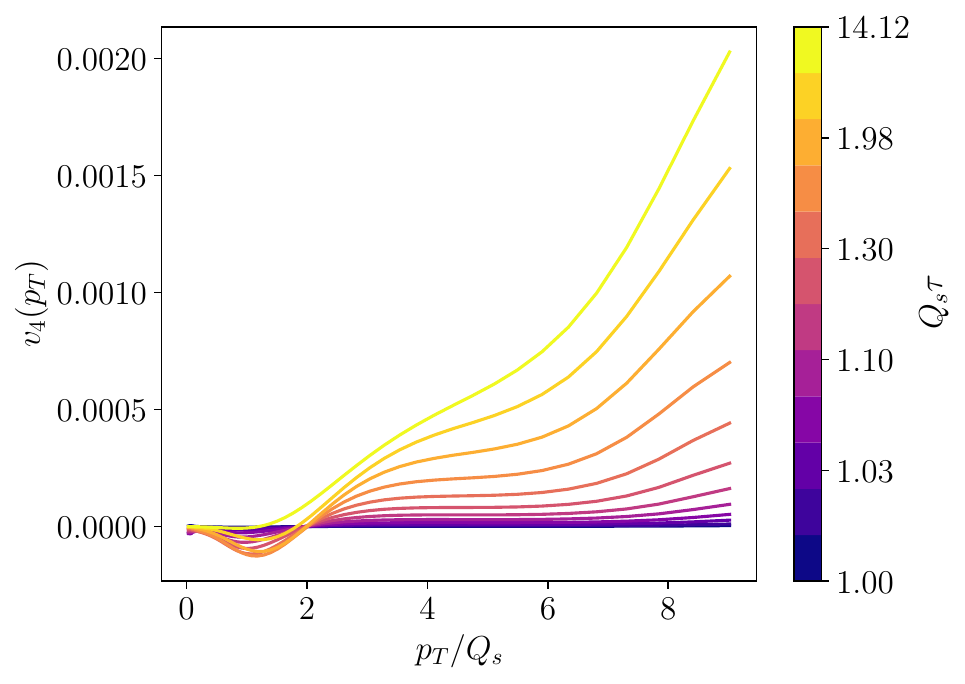}
    \caption{From left to right, values of $v_2$, $v_3$ and $v_4$ as a function of $p_T$ at different times of thermalization, represented with the color gradient shown in the right. Results correspond to an initial condition following Eq.~\eqref{eq:init_vnpt} with $\tilde{v}_2=0.25$ and $\tilde{v}_{n\neq 2}=0$.}
    \label{fig:ptdiff_full}
\end{figure}

The $p_T$-dependent harmonic coefficients are displayed for different times in Fig.~\ref{fig:ptdiff_full}. In the left panel, the evolution of the only non-zero initial coefficient is shown. At the initial time, $v_2$ has a peaked distribution around $p_T\sim Q_s$. By simple inspection of the plot, we observe three different features of the evolution. First, we see that the magnitude of the coefficient decreases with time. This is an expected behavior since the $C_{2\leftrightarrow 2}$ collision kernel isotropizes the system, as it has already been described in the previous section. The second effect that we observe is that the peak is shifted towards higher transverse momenta. Finally, there is a strong enhancement in the value of the $v_2$ at large values of $p_T$. 

\begin{figure}
    \centering
    \includegraphics[trim=0 0 3.2cm 0, clip, height=0.38\linewidth]{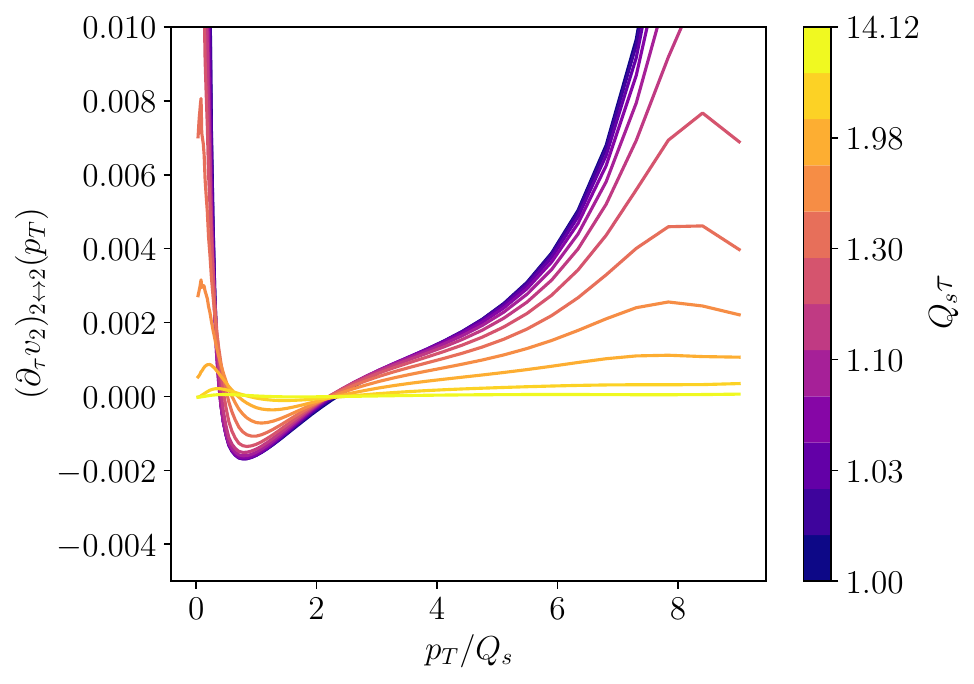}
    \includegraphics[height=0.38\linewidth]{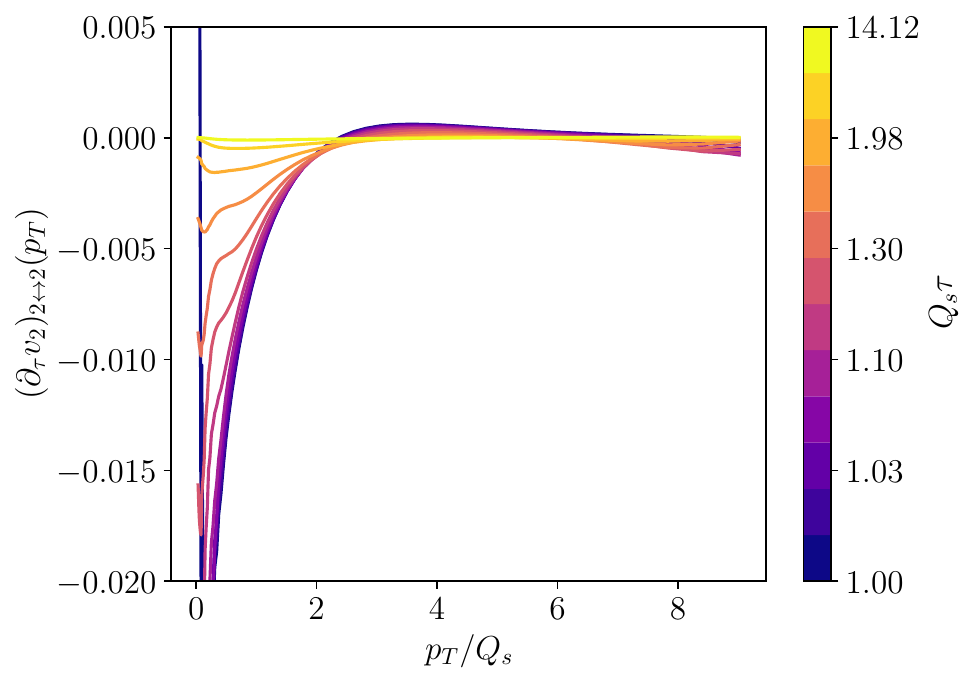}
    \caption{Contribution to the time derivative of the $v_2$ coefficient from the $C_{1\leftrightarrow2}$ (left) and $C_{2\leftrightarrow2}$ (right) collision kernels. Results are taken from the same simulation as in Fig.~\ref{fig:ptdiff_full}.}
    \label{fig:dt_vn}
\end{figure}

If we want to understand the relevant physics contributing to this evolution, we need to examine how the collision kernels modify the value of the $v_2$ coefficient. This can be computed for the present case directly from Eqs.~\eqref{eq:deriv_coef} and~\eqref{eq:deriv_coef_2}. In Fig.~\ref{fig:dt_vn}, we show the results for the $v_2$ contribution from the $2\leftrightarrow2$ and $1\leftrightarrow2$ processes. There are two distinct regimes in the evolution: $p_T \lesssim 2Q_s$ and $p_T \gtrsim 2Q_s$, which we refer to as the IR and UV sectors, respectively.

In the IR sector, the $1\leftrightarrow2$ splitting contribution increases the anisotropy for $p_T \lesssim 0.5~Q_s$. This can be understood as follows. Because of the initial anisotropy, there are more partons with $p_T \sim Q_s$ propagating in the $x$ direction than in the $y$ direction. Since gluon radiation is very efficient at producing a soft thermal bath at early times~\cite{BarreraCabodevila:2022jhi}, the radiated partons inherit the initial anisotropy, which is then transported toward the low-$p_T$ region. The same mechanism also explains the slight decrease of the anisotropy around $p_T \sim Q_s$ induced by inelastic interactions. Enhancing the anisotropy in the deep-IR region requires an energy flow from a higher-momentum region. As discussed above, this energy flow that changes the anisotropy originates primarily from the region with the largest anisotropy, namely $p_T \sim Q_s$. In this case, most of the energy flow happens in the $x$ direction, that is, in the direction in which the initial anisotropy is larger.

If we examine the contribution of elastic interactions in the low-$p_T$ region of phase space, we observe some important differences. First, the $2\leftrightarrow2$ interactions decrease\footnote{At very early times, in the very low-$p_T$ region, the $2\leftrightarrow2$ interaction shows a rapid increase in the anisotropy. This is related to the fact that, at very early times, the soft thermal bath has not yet been built up, and the initial overoccupancy of the system drives the elastic interactions to form a Bose--Einstein condensate, resulting in an energy flow toward the $p \to 0$ region of phase space~\cite{Blaizot:2013lga, Blaizot:2014jna}. Since this formation is associated with a flux of number density in the radial direction, the contribution depends on the azimuthal direction due to the initial anisotropies. Therefore, $v_2$ also increases in the $p_T \to 0$ limit. Once the $1\leftrightarrow2$ processes complete the formation of the soft thermal bath, this behavior disappears.} the anisotropy more rapidly than the $1\leftrightarrow2$ processes can regenerate it after the soft thermal bath has been established. Therefore, the $2\leftrightarrow2$ interactions drive $v_2$ toward zero, efficiently isotropizing the system in the IR region. This explains the mechanism behind the decrease of $v_2$ observed in Fig.~\ref{fig:ptdiff_full}. Moreover, since isotropization is more efficient at lower momenta, it also explains why the peak of the anisotropy shifts toward higher $p_T$ values.

On the other hand, the behaviour of $v_2$ in the UV is drastically different from the one just analyzed. In Fig.~\ref{fig:ptdiff_full}, we observe a pronounced increase in the anisotropy at $p_T \gtrsim 5Q_s$. By examining the contribution of the collision kernels in Fig.~\ref{fig:dt_vn}, it is clear that this behaviour is driven by the $1\leftrightarrow2$ processes, since, in comparison, the elastic contributions barely modify the $v_2$ value for $p_T \gtrsim 2Q_s$. The physical origin of this increase can be traced back to the initial conditions chosen in Eq.~\eqref{eq:InitCond}. The Gaussian tail of the initial distribution implies an extremely low occupancy for $p \gg Q_s$. Thus, although it is a rare process, it is still possible\footnote{It is also necessary, since the tail of the equilibrium distribution is exponential; therefore, high-momentum partons must eventually be produced.} to produce high-$p_T$ gluons via the merging of two lower-momentum gluons. As in the low-$p_T$ case, the merging process is collinear, and the produced partons inherit the anisotropy of the parent gluons. As a consequence, there is an anisotropy flux toward the high-$p_T$ region.

\subsection{Phenomenological study mimicking experimental data}
The initial-state azimuthal anisotropies calculated so far fail to reproduce the measured anisotropies in small collision systems such as $p+\mathrm{Pb}$ (see, \textit{e.g.}~\cite{Schenke:2015aqa}). However, if one extrapolates the pre-hydrodynamic stage of heavy-ion collisions to these types of systems, it is possible that final-state interactions described by kinetic theory modify the initial anisotropies such that a reasonable description of the data becomes possible. In this work, we do not aim to achieve a realistic reproduction of the data, since we neglect spatial gradients, which are expected to play an important role in these systems.

However, we can attempt to mimic the experimental data under the following approximations. First, in the case of proton–lead collisions, the approximation of longitudinal boost invariance is not as good as in lead–lead collisions~\cite{CMS:2017shj, ALICE:2018wma, ALICE:2025luc}. However, the contribution of longitudinal expansion to azimuthal isotropization is not very significant, as discussed in Section~\ref{sec:Azim1}, so it will not affect the present study. Second, the transverse size of the system in this case is of order $\sim 1~\mathrm{fm}$, so if the initially produced system is not strongly anisotropic and we focus on the central region, the effects of spatial gradients are not relevant for $t \ll 1~\mathrm{fm}$. Thus, we can apply our framework to study isotropization.

\begin{figure}
    \centering
    \includegraphics[width=1\linewidth]{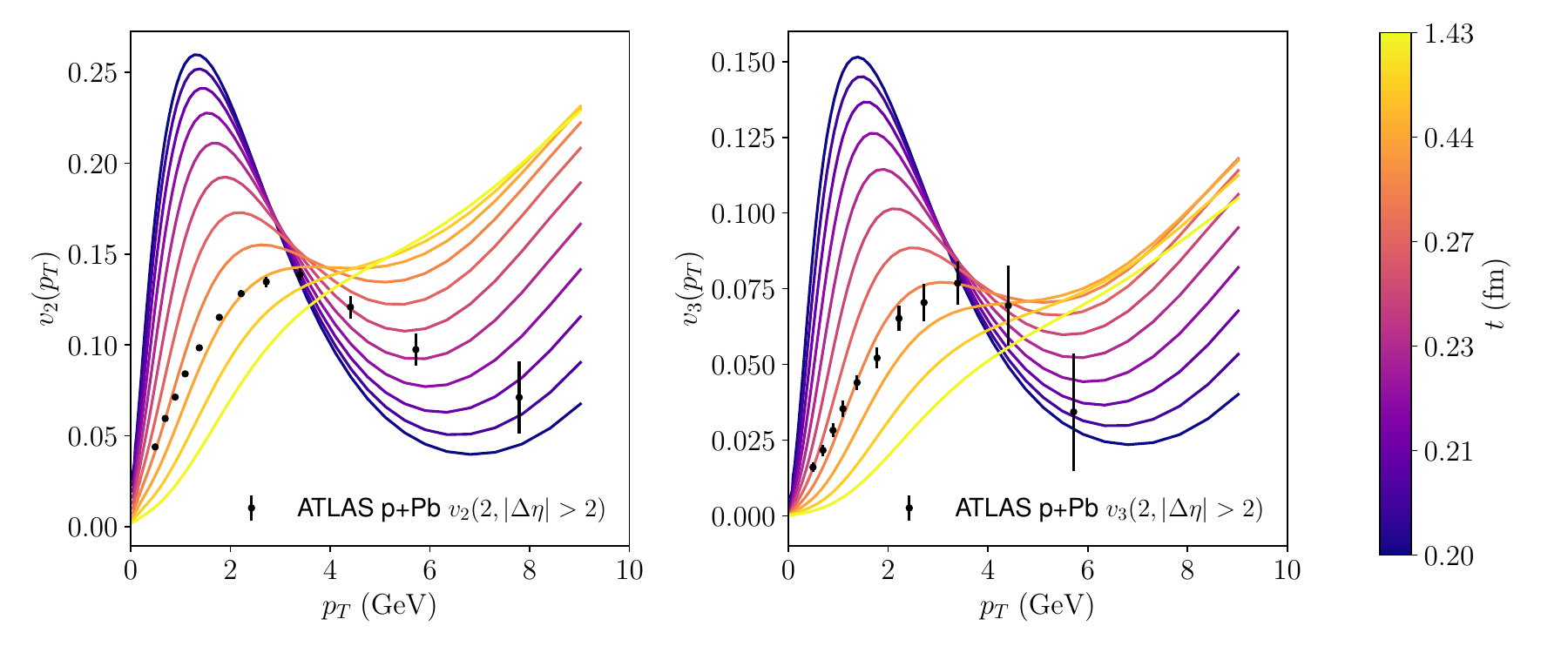}
    \caption{Values of the $v_2$ (left panel) and $v_3$ (right panel) Fourier coefficients as functions of $p_T$ for our system with initial conditions chosen to mimic $p+\mathrm{Pb}$ collisions. The black points correspond to ATLAS data from~\cite{ATLAS:2014qaj}.}
    \label{fig:pPbcomparison}
\end{figure}

The initial conditions are those discussed in the previous section. However, instead of setting the energy density to that corresponding to lead--lead collisions at LHC energies, we parameterize it following Ref.~\cite{Kurkela:2021ctp}. We take the parameter values corresponding to central $p+\mathrm{Pb}$ collisions and assume an initial longitudinal anisotropy of $\xi=5$. For the initial azimuthal anisotropies, we use the decomposition in Eq.~\eqref{eq:init_vnpt} with $\tilde{v}_2=0.75$, $\tilde{v}_3=0.45$, and $\tilde{v}_{n\neq 2,3}=0$, and evolve the Boltzmann equation\footnote{The values chosen for the initial anisotropies are significantly larger than those predicted by current initial-state calculations. However, our goal is not to reproduce the experimental data quantitatively, but rather to demonstrate that the isotropization mechanism discussed above reshapes the initial anisotropies into a form that more closely resembles the experimental observations. For the particular values chosen here, we find a reasonable overlap with the data.}. The results, together with the experimental data from~\cite{ATLAS:2014qaj}, are shown in Fig.~\ref{fig:pPbcomparison}.

As discussed in the previous subsection, the overall magnitude of the peak decreases while it shifts toward higher momenta. We observe that the position of the peak qualitatively matches the data at low momenta at a time of $\sim 0.5~\mathrm{fm}/c$ for both $v_2$ and $v_3$ for $p_T \lesssim 4~\mathrm{GeV}$. For larger momenta, we have already discussed that the large enhancement in the anisotropy is related to the extremely low occupancy of the system in this region, which is due to our parametrization and therefore may not be taken too seriously in this comparison. At this time, the effect of spatial gradients may not be dominant, but it may not be neglected. However, the absence of this effect does not affect the main conclusion that final-state interactions should not be neglected if initial azimuthal momentum anisotropies are produced in the collision.

\begin{figure}
    \centering
    \includegraphics[width=1\linewidth]{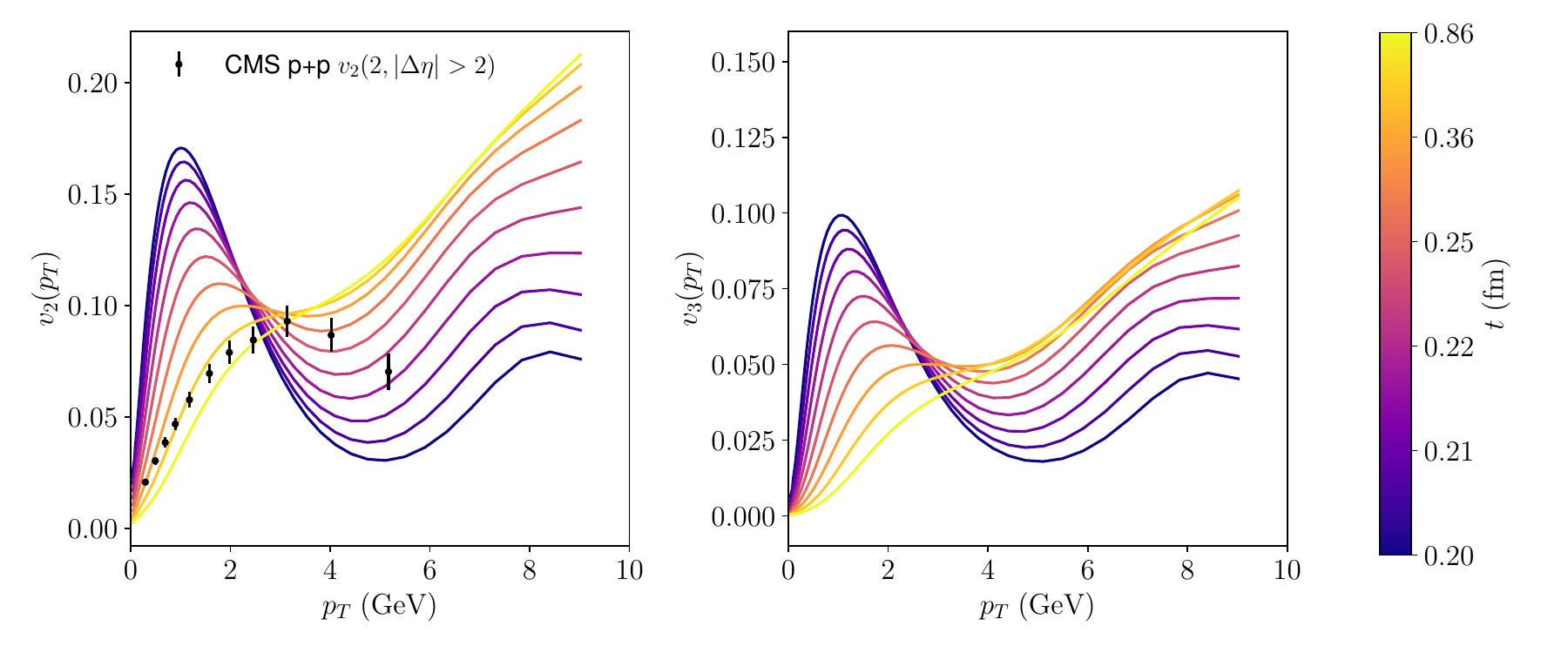}
    \caption{Values of the $v_2$ (left panel) and $v_3$ (right panel) Fourier coefficients as functions of $p_T$ for a system mimicking $p+p$ initial conditions. Black points correspond to CMS data from~\cite{CMS:2016fnw}.}
    \label{fig:ppcomparison}
\end{figure}

We present a similar comparison for $pp$ initial conditions in Fig.~\ref{fig:ppcomparison}. Here, we also initialize the system using the initial energy-density profile from~\cite{Kurkela:2021ctp}. We take $\tau_0 = 0.2~\mathrm{fm}$ and $\xi = 5$, and the initial anisotropies are $v_2 = 0.5$, $v_3 = 0.3$, and $v_{n\neq 2,3} = 0$. In this case, the shift of the peak is not complete until a time $\tau \sim 0.8~\mathrm{fm}$, which is very close to the typical system radius estimated in Ref.~\cite{Kurkela:2021ctp}. Therefore, finite-size effects may be important and could significantly modify the overall behavior.

\begin{figure}
    \centering
    \includegraphics[width=1\linewidth]{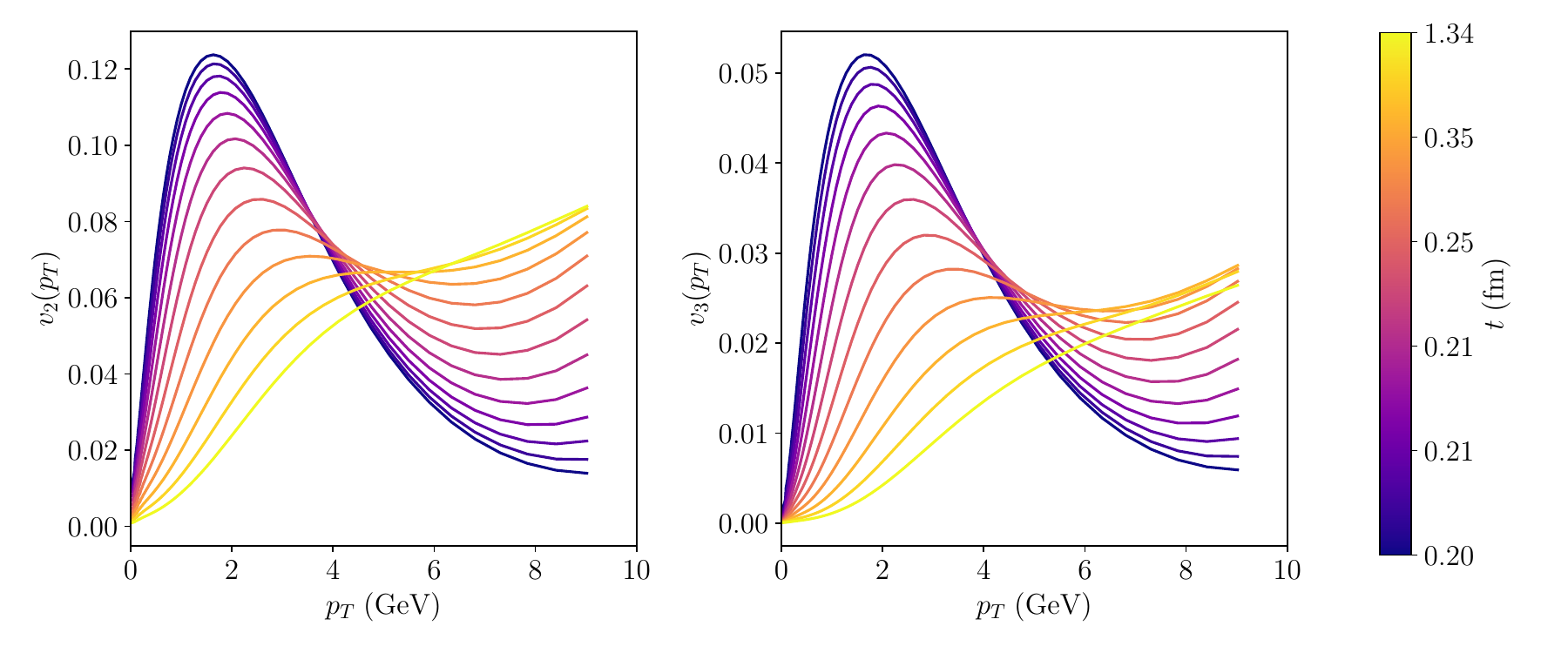}
    \caption{Values of the $v_2$ (left panel) and $v_3$ (right panel) Fourier coefficients as functions of $p_T$ for a system mimicking $O+O$ initial conditions.}
    \label{fig:OO}
\end{figure}

Finally, in Fig.~\ref{fig:OO}, a similar evolution is presented for a system that mimics the initial conditions of oxygen–oxygen collisions. As in the previous cases, we initialise the simulation at $\tau_0 = 0.2~\mathrm{fm}$ and $\xi = 5$. Since there are no data for comparison, we choose an arbitrary amplitude for the initial azimuthal anisotropies, $\tilde{v}_2 = 0.35$ and $\tilde{v}_3 = 0.15$. For this evolution, the anisotropy is very small in the low-momentum sector, where the bulk of the particles resides. The anisotropy remains large in the UV due to the very low occupancy of the system in this region. However, we can conclude that this case, in which the anisotropy in the bulk of the system becomes very small at a time $\tau \sim 1.34~\mathrm{fm}$, shorter than the system radius given by the parametrization~\cite{Kurkela:2021ctp}, indicates that initial-state anisotropies are fully washed out before finite-size effects of the medium become relevant.

\section{Conclusions}\label{sec:Conclusions}

Azimuthal correlations are a strong signal of collectivity and were historically used as evidence for the existence of a thermal Quark-Gluon Plasma, as they provide a smoking-gun signature of a relativistic hydrodynamic description. However, the observation of similar signals in small collision systems, such as proton–proton or proton–lead collisions, where the applicability of hydrodynamics is under debate, raised the question of whether non-hydrodynamic mechanisms could also generate such anisotropies. Various works have proposed that azimuthal correlations may already arise from the initial state of the collision as an alternative to the hydrodynamic explanation.

In this paper, we study for the first time how these initial anisotropies relax over time due to final-state interactions, and what their possible effects are on the thermalization process. To this end, we use the Boltzmann Equation in the Diffusion Approximation (BEDA), which has previously been applied to the study of thermalization in longitudinally expanding, boost-invariant, and azimuthally symmetric plasmas. We extend it to the initial conditions of a purely gluonic system containing azimuthal anisotropies encoded in a few harmonic coefficients and study their time evolution. Since the distribution function evolved in the BEDA is directly related to the particle spectra measured experimentally, it can be connected to the azimuthal anisotropies observed in experimental data.

First, we explore the relaxation of the fully integrated harmonic coefficients. This simplified picture neglects any momentum dependence that would arise from a full Fourier decomposition and integrates it out. This allows us to study azimuthal isotropization straightforwardly. In particular, we observe a hierarchy in the isotropization dynamics, where higher-order harmonics (larger $n$) relax faster than lower-order ones. In systems where quark production is allowed, the isotropization is slower, since fermionic interactions are generally less efficient than gauge-field interactions in QCD. This difference is smaller at higher harmonic order, since the anisotropies decay so rapidly that by the time quark production becomes relevant, the gluonic system is already close to azimuthal isotropy.

Secondly, we study in more detail the momentum dependence of the Fourier coefficients. In general, $v_n = v_n(p_T)$, as predicted in initial-state calculations of azimuthal anisotropies. These calculations typically predict a peaked structure in $p_T$ around the saturation momentum $Q_s$, which we therefore model into our initial conditions. In this case, we observe that the combined effect of $2\leftrightarrow2$ interactions in the infrared and $1\leftrightarrow2$ processes in the ultraviolet shifts the peak towards higher momenta. This is an important observation, since in small collision systems the peak of $v_n(p_T)$ usually appears at larger momenta than predicted by initial-state calculations alone.

Motivated by this result, we conclude by performing a phenomenological study using initial energy densities consistent with those expected in small collision systems and compare our results with experimental data. In particular, we consider proton–proton and proton–lead collisions and perform a similar study for oxygen–oxygen collisions, for which no data are yet available. We fix an overall normalization for the initial anisotropy amplitude and observe that the peak position is shifted in the direction suggested by the experimental measurements. This indicates that final-state interactions can reproduce key features of the data if the initial anisotropy is sufficiently large, without requiring transverse collective flow, which is a central ingredient in hydrodynamic descriptions. However, the effects of spatial gradients, which are neglected in our present setup and would account for radial expansion, may have important consequences and could significantly modify these conclusions. We leave the inclusion of transverse spatial dependence for future work. In the case of oxygen–oxygen collisions, we find that most of the initial-state anisotropy is damped before the finite transverse size effects become relevant, suggesting that it may play a subleading role in the experimental observations.

\appendix

\section{Fully integrated \texorpdfstring{$v_n$}{vn} variation from other collision kernels}\label{sec:app_others}

The contribution from the source term is trivial when we consider the harmonic coefficient of the full system, that is, the one that we study in this work. In this case, it is straightforward to check that
\begin{eqnarray}
    \nu_gS_g + \nu_q N_f (S_q + S_{\bar{q}}) = 0~,   
\end{eqnarray}
so there will be no variation of the $v_n$ coefficients. 

However, this is not that simple for the $1\leftrightarrow2$ splitting/merging processes. Let us expand in detail Eq.~\eqref{eq:inel_kern} for the 2 relevant species that play a role in our work (assume quark-antiquark symmetry),
\begin{align}
\begin{split}
        C_{1\leftrightarrow2}^g (\mathbf{p}) = \int_0^1 \frac{dx}{x^3}\left[ C^g_{gg} + \frac{N_f}{C_F} C^q_{gq} \right]\left(\frac{\mathbf{p}}{x}, \mathbf{p}, \frac{\mathbf{p}\bar{x}}{x}\right) \\- \int_0^1 dx \left[ \frac{1}{2}C^g_{gg} + N_f C^g_{qq}\right] (\mathbf{p}, \mathbf{p}x, \mathbf{p}\bar{x})~,
\end{split}
\end{align}

\begin{eqnarray}
        C_{1\leftrightarrow2}^q (\mathbf{p}) = \int_0^1 \frac{dx}{x^3}\left[ C^q_{qg} + 2C_F C^g_{qq} \right]\left(\frac{\mathbf{p}}{x}, \mathbf{p}, \frac{\mathbf{p}\bar{x}}{x}\right) - \int_0^1 dx C^q_{qg}(\mathbf{p}, \mathbf{p}x, \mathbf{p}\bar{x})~.
\end{eqnarray}
Because we are interested in exploring integrals of the type
\begin{eqnarray}
    \int d^3\mathbf{p}\left( \nu_g C_{1\leftrightarrow2}^g (\mathbf{p}) + 2\nu_q N_f C_{1\leftrightarrow2}^q (\mathbf{p})\right) h(\phi)~,
\end{eqnarray}
we can get rid of the $1/x^3$ factor in the collision kernels and obtain that
\begin{eqnarray}
    \int d^3 \mathbf{p}~C^{\mathrm{all}}_{1\leftrightarrow2} (\mathbf{p}) h(\phi) = \int d^3\mathbf{p} dx\left[\frac{\nu_g}{2} C^g_{gg} + 2\nu_q N_f C^q_{qg} + \nu_g N_f C^g_{qq} \right](\mathbf{p}, \mathbf{p}x, \mathbf{p}\bar{x}) h(\phi)
\end{eqnarray}

Remember that our goal is to explore the evolution of the harmonic coefficients according to Eq.~\eqref{eq:deriv_coef}. For the $C^a_{bc}$, the only relevant part in the evolution that actually matters is that with information about the azimuthal profile, that is, the distribution functions. Thus, let us forget about the splitting rates and the integral in $x$, which is not needed for a qualitative analysis, and focus on the statistical term. In QCD, the only terms in the statistical term for the $1\leftrightarrow2$ process are the linear and quadratic ones,
\begin{eqnarray}
    \mathcal{F}^a_{bc}(\mathbf{p}, \mathbf{k}, \mathbf{q}) = f^a_\mathbf{p} + \epsilon_b f^a_\mathbf{p}f^b_\mathbf{k} + \epsilon_c f^a_\mathbf{p}f^c_\mathbf{q} - f^b_\mathbf{k}f^c_\mathbf{q}~.
\end{eqnarray}
By inserting the ansatz~\eqref{eq:ansatz}, we can rewrite this expression as
\begin{align}
\begin{split}
    \mathcal{F}^a_{bc}(\mathbf{p}, \mathbf{k}, \mathbf{q}) = \alpha^{a\to bc}(\mathbf{p}, \mathbf{k}, \mathbf{q}) + 2\sum_n b^{a\to bc}_m(\mathbf{p}, \mathbf{k}, \mathbf{q}) \cos(m\phi) \\+ 4\sum_{ml} c^{a\to bc}_{ml}(\mathbf{p}, \mathbf{k}, \mathbf{q}) \cos(m\phi) \cos(l\phi)
\end{split}
\end{align}
where we have defined
\begin{align}
    \alpha^{a\to bc}(\mathbf{p}, \mathbf{k}, \mathbf{q}) &\equiv \tilde{f}^a_{\mathbf{p}} + \epsilon_b \tilde{f}^a_{\mathbf{p}}\tilde{f}^b_{\mathbf{k}} + \epsilon_c \tilde{f}^a_{\mathbf{p}} \tilde{f}^c_{\mathbf{q}} - \tilde{f}^b_{\mathbf{k}}\tilde{f}^c_{\mathbf{q}} \\
    b^{a\to bc}_m(\mathbf{p}, \mathbf{k}, \mathbf{q}) &\equiv \tilde{f}^a_{\mathbf{p}}v_m^a + \epsilon_b \tilde{f}^a_{\mathbf{p}}\tilde{f}^b_{\mathbf{k}}(v_m^a + v_m^b) + \epsilon_c \tilde{f}^a_{\mathbf{p}} \tilde{f}^c_{\mathbf{q}}(v_m^a + v_m^c) - \tilde{f}^b_{\mathbf{k}}\tilde{f}^c_{\mathbf{q}}(v_m^b + v_m^c) \\
    c^{a\to bc}_{ml}(\mathbf{p}, \mathbf{k}, \mathbf{q}) &\equiv \epsilon_b \tilde{f}^a_{\mathbf{p}}\tilde{f}^b_{\mathbf{k}} v_m^a v_l^b + \epsilon_c \tilde{f}^a_{\mathbf{p}} \tilde{f}^c_{\mathbf{q}}v_m^a v_l^c - \tilde{f}^b_{\mathbf{k}}\tilde{f}^c_{\mathbf{q}}v_m^b v_l^c
\end{align}

Now we can finally explore how the $1\leftrightarrow2$ collision kernel changes the $v_n$ coefficients. In Eq.~\eqref{eq:deriv_coef}, there are three distinct terms. First, the term proportional to $\dot{\psi}$ will cancel because with the ansatz~\eqref{eq:ansatz}
\begin{eqnarray}
    \int d^3\mathbf{p} \sin(n\phi) f(\mathbf{p}) = 0~.
\end{eqnarray}
Here, we set for convenience the event plane angle to $\psi=0$. Besides this, there is another term that is straightforward to compute
\begin{eqnarray}
    \frac{\dot{N}}{N^2}\int \frac{d^3\mathbf{p}}{(2\pi)^3}\cos (n\phi)f = \frac{\dot{N}}{N}v_n~.
\end{eqnarray}
Thus, this is a relaxation-like term.

The last term has a far from trivial expression, but with the notation that we have introduced previously can be summarized into the following expression
\begin{align}
\begin{split}
    \frac{1}{N}\int \frac{d^3\mathbf{p}}{(2\pi)^3} \cos (n\phi) C = \int \frac{dp p^2 d\cos\theta}{N(2\pi)^3} dx \left[\frac{\nu_g}{2} \frac{d^2I_{g\leftrightarrow gg}}{dx dt}\left(b_n^{g\to gg} + \sum_{ml} c_{ml}^{g\to gg} \Omega_{nml}\right)(\mathbf{p}, \mathbf{p}x, \mathbf{p}\bar{x}) \right. \\ 
    \left. + 2\nu_q N_f \frac{d^2I_{q\leftrightarrow qg}}{dx dt} \left(b_n^{q\to qg} + \sum_{ml} c_{ml}^{q\to qg}\Omega_{nml} \right)(\mathbf{p}, \mathbf{p}x, \mathbf{p}\bar{x}) \right. \\ 
    \left. +\nu_g N_f \frac{d^2I_{g\leftrightarrow qq}}{dx dt} \left(b_n^{g\to qq} + \sum_{ml} c_{ml}^{g\to qq}\Omega_{nml} \right)(\mathbf{p}, \mathbf{p}x, \mathbf{p}\bar{x}) \right]~,
\end{split}
\end{align}
and we have defined
\begin{eqnarray}
    \Omega_{nml} \equiv (\delta_{n,m+l} + \delta_{m,n+l} + \delta_{l,n+m})~.
\end{eqnarray}
This is the same term that showed up in the $2\leftrightarrow2$ case. Thus, it will generate higher harmonics similarly.

At the end of the day, the evolution equation for the harmonic coefficient $v_n$ due to the $1\leftrightarrow2$ interaction is
\begin{align}
\begin{split}
    (\partial_\tau v_n)_{1\leftrightarrow2} = \frac{\dot{N}}{N}v_n - \int \frac{dp p^2 d\cos\theta}{N(2\pi)^3} dx \left[\frac{\nu_g}{2} \frac{d^2I_{g\leftrightarrow gg}}{dx dt}\left(b_n^{g\to gg} + \sum_{ml} c_{ml}^{g\to gg} \Omega_{nml}\right)(\mathbf{p}, \mathbf{p}x, \mathbf{p}\bar{x}) \right. \\ 
    \left. + 2\nu_q N_f \frac{d^2I_{q\leftrightarrow qg}}{dx dt} \left(b_n^{q\to qg} + \sum_{ml} c_{ml}^{q\to qg}\Omega_{nml} \right)(\mathbf{p}, \mathbf{p}x, \mathbf{p}\bar{x}) \right. \\ 
    \left. +\nu_g N_f \frac{d^2I_{g\leftrightarrow qq}}{dx dt} \left(b_n^{g\to qq} + \sum_{ml} c_{ml}^{g\to qq}\Omega_{nml} \right)(\mathbf{p}, \mathbf{p}x, \mathbf{p}\bar{x}) \right]~.
\end{split}
\end{align}
The first term is a relaxation term, as well as the terms inside the integral included in $b^{a\to bc}$. However, notice that these relaxation terms are $\propto v_n$, while the one in the $2\leftrightarrow2$ is $\propto v_n^2$. The term that generates higher harmonics is included in $c^{a\to bc}$.

\section{Numerical implementation of the BEDA code}\label{sec:app_numerics}

The numerical results presented in this work have been obtained with a solver of the BEDA designed to exploit the parallelizability of the calculation of the collision kernels on GPUs. In this section, we describe the algorithm used to solve Eq.~\eqref{eq:BE}. In the following subsections, we explain in detail how each of the terms in the Boltzmann Equation is computed, and finally, we describe the overall parallelized algorithm.

It is useful to introduce some conventions for the following discussion. First, we can rewrite Eq.~\eqref{eq:BE} as 
\begin{eqnarray}\label{eq:BE_num}
    \partial_\tau (pf^a)(\tau;\mathbf{p}) = C_{exp}[f] + C_{1\leftrightarrow2}[f] + C_{2\leftrightarrow2}[f]~.
\end{eqnarray}
In the following, we explain how each of the three terms on the right-hand side is computed in our algorithm. By construction, our algorithm will evolve $pf$ instead of $f$. Secondly, we will assume that the distribution lives in a grid in spherical coordinates for momentum of size $(N_p, N_\theta, N_\phi)$, define by $\{p_i\}_{i=0}^{N_p}$, $\{\cos\theta_j\}_{j=0}^{N_p}$ and $\{\phi_k\}_{k=0}^{N_\phi}$. In general, the algorithm works for any arbitrary grid, but the one used in this work corresponds to a logarithmic grid in $p$ and linear grids in $\cos\theta$ and $\phi$. To avoid evaluation of the $C_{2\leftrightarrow2}$ at singular points, we choose a infrared cut-off for $p$, $p_0=p_{min}\neq 0$, and, similarly, $\cos\theta_0=-1+\Delta_\theta$ and $\cos\theta_{N_\theta}=1-\Delta_\theta$. It is also very convenient to introduce the corresponding special grids for each of the spherical coordinates as $x_{s,i} = (x_i+x_{i+1}) / 2$, with sensible values for the boundaries\footnote{Sensible values for the boundaries are $p_{s,0}=0$, $p_{s,N_p+1}=2p_{N_p}-p_{N_p-1}$, $\cos\theta_{s,0}=-1$, $\cos\theta_{s,N_\theta+1}=-1$, $\phi_{s,0}=\phi_{s,N_\phi+1}$.}.

The three terms on the right-hand side of Eq.~\eqref{eq:BE_num} exhibit conservation of some integral moments of the distribution function, such as number and energy density. In our code, we make energy density exactly conserved under the discretization in terms of wedge functions~\cite{Kurkela:2012hp, Du:2020dvp}
\begin{eqnarray}\label{eq:discrete_e}
    \epsilon = \int \frac{d^3\mathrm{p}}{(2\pi)^3} p f(\mathbf{p}) \equiv \sum_{ijk}\int \frac{d^3\mathrm{p}}{(2\pi)^3} p~w_{ijk}(\mathbf{p}) f(\mathrm{p})~,
\end{eqnarray}
with
\begin{eqnarray}
    w_{ijk}(\mathrm{p}) \equiv w_{i}(p) w_{j}(\cos\theta) w_{k}(\phi)~,
\end{eqnarray}
and
\begin{equation}\label{eq:wedges}
    w_i(x) = \begin{cases}
\frac{x_i-x}{x_{i}-x_{i-1}} & \text{if } x\in[x_{i-1},x_i] \\
\frac{x_{i+1}-x}{x_{i+1}-x_{i}} & \text{if } x\in[x_{i},x_{i+1}] \\
0 & \text{otherwise}
\end{cases}~.
\end{equation}
Similarly, number density is defined as 
\begin{eqnarray}\label{eq:discrete_n}
    n \equiv \sum_{ijk}\int \frac{d^3\mathrm{p}}{(2\pi)^3} w_{ijk}(\mathbf{p}) f(\mathrm{p})~.
\end{eqnarray}

\subsection{The expansion term}

The expansion term in Eq.~\eqref{eq:BE_num} in spherical coordinates can be written as a conservation-like equation as
\begin{eqnarray}
    C_{exp}[f] = \frac{p}{\tau} \left( \frac{\cos^2 \theta}{p^2}  \frac{\partial (p^3 f)}{\partial p} + \frac{\partial\left[ (\cos\theta - \cos^3 \theta)f \right]}{\partial \cos \theta}\right) - \frac{pf}{\tau}.
\end{eqnarray}
Analytically, this term implies an exact solution for the number density $n(t) = n(t_0)/t$, which can be made exact for the definition~\eqref{eq:discrete_n} under the discretization scheme
\begin{align}
\begin{split}
    C_{exp}^{ijk}[f] \approx \frac{p_i}{\tau} & \left( \frac{\cos^2\theta_j}{p_i^2} \frac{(p^3f)_{i_s+1ij} - (p^3f)_{i_sij}}{\Delta p_{i_s}} + \right. \\ &~~\left. \frac{((\cos\theta - \cos^3 \theta)f)_{ij_s+1k} - ((\cos\theta - \cos^3 \theta)f)_{ij_sk}}{\Delta \cos\theta_{j_s}}\right) - \frac{(pf)_{ijk}}{\tau}~,
\end{split}
\end{align}
where the subscript $s$ resembles the evaluation of the distribution function on the special grid. This evaluation is obtained by linear interpolation of the points that rest on the regular grid.

\subsection{The \texorpdfstring{$C_{1\leftrightarrow2}$}{C12} collision integral}

As in Refs.~\cite{AbraaoYork:2014hbk, Du:2020dvp}, we implement the $1\leftrightarrow2$ collision kernels in such a way that energy density of Eq.~\eqref{eq:discrete_e} is exactly conserved with a finite elements method. To see that, let us start with the general expression for the contribution to the arbitrary $a\leftrightarrow bc$ process. In this case, the collision kernels affecting each of the 3 involved partons are
\begin{eqnarray}
    C^{a,a\leftrightarrow bc}_{1\leftrightarrow2}[f] (p) &= - \displaystyle\int_0^{\frac{1}{2}} dx C^{a}_{bc}(p, px, p\bar{x}) ~, \\
    C^{b,a\leftrightarrow bc}_{1\leftrightarrow2}[f] (p) &= \displaystyle\int_0^\frac{1}{2}\frac{dx}{x^3} C^a_{bc}(p/x, p, p\bar{x}/x)~,\\
    C^{c,a\leftrightarrow bc}_{1\leftrightarrow2}[f] (p) &= \displaystyle\int_0^\frac{1}{2}\frac{dx}{\bar{x}^3} C^a_{bc}(p/\bar{x}, px/\bar{x}, p)~,
\end{eqnarray}
where, for the sake of simplicity, we define $\bar{x}\equiv1-x$. The coefficient $C^a_{bc}$ is the same as defined in Eq.~\eqref{eq:inel_kern2}, and it will be irrelevant in the following discussion. From this point, one can show energy conservation analytically,
\begin{eqnarray}
    \int_\mathbf{p} p \left[C^{a,a\leftrightarrow bc}_{1\leftrightarrow2}[f] (p) + C^{b,a\leftrightarrow bc}_{1\leftrightarrow2}[f] (p) + C^{c,a\leftrightarrow bc}_{1\leftrightarrow2}[f] (p) \right] = 0~.
\end{eqnarray}
As they are written, the terms in the equations above represent the time variation in the distribution function $f$ at the momentum $p$. We can rewrite the integrand for the energy flux of the element $i$ for each of the three particles involved in the $a\leftrightarrow bc$ process\footnote{Since the splitting is collinear, let us assume that the angular variables have been integrated out.}. 
\begin{eqnarray}
    \tilde{C}^a_{i}(x) &= - \displaystyle\int_\mathbf{p} p~w_{i}(p) C^{a}_{bc}(p, px, p\bar{x})~,\\
    \tilde{C}^b_{i}(x) &= \displaystyle\int_\mathbf{p}(px) w_{i}(px) C^a_{bc}(p,px,p\bar{x})~,\\
    \tilde{C}^c_{i}(x) &= \displaystyle\int_\mathbf{p}(p\bar{x}) w_{i}(p\bar{x}) C^a_{bc}(p,px,p\bar{x})~.
\end{eqnarray}
In the last two expressions, we performed a change of variables to remove the $1/x^3$ and $1/\bar{x}^3$ contributions in the integrands. Notice now that, since the base functions~\eqref{eq:wedges} obey the relation $\sum_i w_i(p) = 1$, it is clear that this discretization will conserve energy as given by Eq.~\eqref{eq:discrete_e}.

The $\tilde{C}^a_i$ kernels written above determine the evolution of the element of the distribution function
\begin{eqnarray}
    \epsilon_{ijk} \equiv \displaystyle\int_\mathbf{p} w_{ijk}(p,\cos\theta, \phi) pf~.
\end{eqnarray}
However, this expression can be inverted approximately as
\begin{eqnarray}
    (pf)_{ijk} \approx \frac{\epsilon_{ijk}}{\Delta V_{ijk}}~, \quad \Delta V_{ijk} \equiv \frac{p_i^2 \Delta p_{i_s} \Delta \cos\theta_{j_s} \Delta \phi_{k_s}}{(2\pi)^2}~.
\end{eqnarray}
Thus, we can compute the time derivative of the distribution function for each parton involved in the process as
\begin{eqnarray}
    \left[\partial_t (pf)^a_{ijk}\right]_{a\leftrightarrow bc} \approx \frac{1}{\Delta V_{ijk}} \sum_n \Delta x_n \tilde{C}^a_i(x_n)~, \\
    \left[\partial_t (pf)^b_{ijk}\right]_{a\leftrightarrow bc} \approx \frac{1}{\Delta V_{ijk}} \sum_n \Delta x_n \tilde{C}^b_i(x_n)~, \\
    \left[\partial_t (pf)^c_{ijk}\right]_{a\leftrightarrow bc} \approx \frac{1}{\Delta V_{ijk}} \sum_n \Delta x_n \tilde{C}^c_i(x_n)~,
\end{eqnarray}
such that energy will be exactly conserved. This procedure can be repeated for each of the possible $a\leftrightarrow bc$ processes allowed by the QCD interaction vertices, so it calculates all of the $1\leftrightarrow2$ collision kernel contributions.

\subsection{The \texorpdfstring{$C_{2\leftrightarrow2}$}{C22} collision integral}\label{sec:C22}

The $2\leftrightarrow2$ collision kernel corresponds to a Fokker-Planck-like equation plus a source term. The integration of this term in the algorithm corresponds to an implicit-explicit method that can be shown to conserve exactly the number density, as one can check analytically from Eq.~\eqref{eq:C2to2}. On the other hand, energy density, which can also be shown to be conserved analytically, is not exactly conserved in our algorithm, but we introduce some modifications that improve its conservation drastically. The integration method we use takes care of the diffusive part implicitly with a Douglas-Gunn scheme, while the terms without derivatives, or with derivatives of order one, are computed explicitly. With this, we get rid of the strongest source of instabilities for a naive finite difference method related to the second-order derivatives that impose a CFL condition $\Delta t \propto 1 / \Delta x^2$.

Let us write the Boltzmann equation we want to solve as
\begin{eqnarray}
    \partial_\tau f = \frac{\hat{q}}{4} \nabla_{\mathbf{p}}^2 f + \tilde{\mathcal{L}}[f],
\end{eqnarray}
where $\tilde{\mathcal{L}}[f]$ involves the rest of the terms that are not diffusive. By expanding the Laplacian,
\begin{align}
    \partial_\tau f = \frac{\hat{q}}{4} \left[ \frac{1}{p^2}\frac{\partial}{\partial p}\left(p^2\frac{\partial f(\vec{p},t)}{\partial p}\right)+\frac{1}{p^2}\frac{\partial }{\partial\cos\theta}\left(\sin^2\theta\frac{\partial f(\vec{p},t)}{\partial\cos\theta}\right)+\frac{1}{p^2\sin^2\theta}\frac{\partial^2 f(\vec{p},t)}{\partial^2\phi} \right] + \tilde{\mathcal{L}}[f].
\end{align}
For convenience, since we will evolve $pf$ instead of $f$, we can write the previous equation as
\begin{align}
\label{eq:preDG}
    \partial_\tau(pf) = \frac{\hat{q}}{4} \left[ \frac{p}{p^2}\frac{\partial}{\partial p}\left(p\frac{\partial (pf)}{\partial p}\right)+\frac{1}{p^2}\frac{\partial }{\partial\cos\theta}\left(\sin^2\theta\frac{\partial (pf)}{\partial\cos\theta}\right)+\frac{1}{p^2\sin^2\theta}\frac{\partial^2 (pf)}{\partial^2\phi} \right] + \mathcal{L}[f],
\end{align}
where we have defined
\begin{eqnarray}
    \mathcal{L}[f] \equiv \tilde{\mathcal{L}}[f] - \frac{\hat{q}}{4} \frac{p}{p^2}\frac{\partial(pf)}{\partial p}~.
\end{eqnarray}
Let us also rewrite Eq.~\eqref{eq:preDG} as
\begin{eqnarray}
    \partial_\tau(pf) = \frac{\hat{q}}{4} \left[ \frac{p}{p^2}\frac{\partial J_p}{\partial p}+\frac{1}{p^2}\frac{\partial J_\theta }{\partial\cos\theta}+\frac{1}{p^2\sin^2\theta}\frac{\partial J_\phi}{\partial\phi} \right] + \mathcal{L}[f],
\end{eqnarray}
and apply the DG method to obtain the next step for the distribution function while treating $\mathcal{L}$ implicitly. The currents $J_p$, $J_\theta$ and $J_\phi$ will be computed with finite differences and evaluated in the $s$-grid. For completeness, let us write them below:
\begin{eqnarray}
    J_{p,s}^{ijk} &=& p_{i_s} \frac{(pf)_{ijk} - (pf)_{i-1jk}}{\Delta p_{i}}~, \\
    J_{\theta, s}^{ijk} &=& (1-\cos^2 \theta_{j_s})\frac{(pf)_{ijk} - (pf)_{ij-1k}}{\Delta \cos \theta_{j}}~, \\
    J_{\phi, s}^{ijk} &=& \frac{(pf)_{ijk} - (pf)_{ijk-1}}{\Delta \phi_{k}}~.
\end{eqnarray}
Notice that the appropriate boundary conditions must be applied. Since the second-order derivatives are deeply related to the $C_{2\leftrightarrow2}$ term, it is sensible to impose them such that the number conservation is granted. These corresponds to $J_{p,s}^{0jk}=J_{p,s}^{N_pjk}=0$, $J_{\theta,s}^{i0k}=J_{\theta,s}^{iN_\theta k}=0$ and $J_{\phi,s}^{ij0}=J_{\phi,_s}^{ijN_\phi}$.

The Douglas-Gunn method computes the distribution function at time $t+\Delta t$, $f^{n+1}$, from the value of the distribution at $t$, $f^n$, in three different steps.
\begin{enumerate}
    \item First, we compute the value of $(pf)^*$ given by the following equation where we make explicit the times
    \begin{eqnarray}
        \frac{(pf)^{*} - (pf)^n}{\Delta \tau}\Bigg |_{ijk} = \frac{\hat{q}}{4} && \left[ \frac{1}{p} \frac{1}{2}\left(\frac{J_{ps,i+1}^{*}-J_{ps,i}^{*}}{\Delta p_{s,i}} + \frac{J_{ps,i+1}^{n}-J_{ps,i}^{n}}{\Delta p_{s,i}} \right)+ \right.\\
        && \left. \frac{1}{p^2}\frac{J_{\theta s,j+1}^n - J_{\theta s,j}^n }{\Delta \cos\theta_{s,j}}+ \right. \\
        && \left. \frac{1}{p^2\sin^2\theta}\frac{J_{\phi s, k+1}^n - J_{\phi s, k}^n}{\Delta \phi_{s,k}} \right] + \mathcal{L}[f]~.
    \end{eqnarray}
    To avoid writing more indices, the currents only include the one exclusively related to the ones related with the derivative, and the rest must be interpreted as the one for the position we are computing.

    \item The next step computes $f^{**}$ as
    \begin{eqnarray}
        \frac{(pf)^{**}-(pf)^*}{\Delta \tau}\Bigg |_{ijk} = \frac{\hat{q}}{4} \frac{1}{p^2} \frac{1}{2} \left[ \frac{J_{\theta s,j+1}^{**} - J_{\theta s,j}^{**} }{\Delta \cos\theta_{s,j}} - \frac{J_{\theta s,j+1}^* - J_{\theta s,j}^* }{\Delta \cos\theta_{s,j}} \right]~.
    \end{eqnarray}

    \item Finally, we compute $f^{n+1}$ as 
    \begin{eqnarray}
        \frac{(pf)^{n+1}-(pf)^{**}}{\Delta \tau}\Bigg |_{ijk} = \frac{\hat{q}}{4} \frac{1}{p^2} \frac{1}{2} \left[ \frac{J_{\phi s,k+1}^{n+1} - J_{\phi s,k}^{n+1} }{\Delta \phi{s,k}} - \frac{J_{\phi s,k+1}^{**} - J_{\phi s,k}^{**} }{\Delta \phi_{s,k}} \right]~.
    \end{eqnarray}
\end{enumerate}
This method implicitly solves only the term regarding the diffusion part, as required. All the terms included in $\mathcal{L}$ are computed explicitly in the first step. 

The first two steps of the Douglas-Gunn algorithm require inverting a tridiagonal matrix, for which we use a Thomas algorithm. For the $\phi$ derivative, because of the periodic boundary conditions, the matrix we need to invert is not tridiagonal, since it has non-zero values in the corner elements of the secondary diagonal. In this case, we use the Sherman-Morrison formula to reduce the problem to solve to another one suitable for the Thomas algorithm.

\subsubsection{The \texorpdfstring{$\mathcal{L}$}{L} term}

The term that is integrated explicitly, $\mathcal{L}$, includes all the contributions to the $2\leftrightarrow2$ collision kernel that are not second-order derivatives. Also, to do a consistent integration (which also produces more stable results), we also include here the conversion term, the $1\leftrightarrow2$ collision kernel, and the expansion term. That is,
\begin{eqnarray}
    \mathcal{L}_{ijk} \equiv \frac{\hat{q}}{4p}\frac{\partial}{\partial p}\left[ \frac{1}{T_*} pf ( p+pf) -pf \right]_{ijk} + p\mathcal{S}_{ijk} + p_iC^{1\leftrightarrow2}_{ijk} + p_iC^{exp}_{ijk} - \\
    -\frac{\hat{q}_{eq}}{4p}\frac{\partial}{\partial p}\left[ \frac{1}{T_{eq}} pf_{eq} ( p+pf_{eq}) -pf_{eq} +p \frac{\partial(pf_{eq})}{\partial p}\right]_{ijk}~.
\end{eqnarray}
This definition includes an extra term, which evaluates the $C_{2\leftrightarrow2}$ kernel for a function in equilibrium, $f_{eq}$, which corresponds to the equilibrium distribution for a system with the same energy density as the one that we are evaluating. If we compute this term analytically, it will be exactly zero. However, since we are using a finite difference scheme to compute the collision kernel, this is not true, and some numerical errors will arise. We subtract this term because it enforces the implementation of the thermal fixed point correctly. We also noticed that it significantly improves the conservation of the energy density in the elastic collision kernel, which is not implemented by construction as it is for the $C_{1\leftrightarrow2}$. 

In general, for a system of gluons, quarks, and antiquarks, the evaluation of the parameters at equilibrium involves solving a system of three nonlinear equations. Let us describe how we compute them in the code. We assume that, for the most general system, the distributions at equilibrium for each of the particles are
\begin{eqnarray}
    f_{eq} = \frac{1}{e^{\frac{p-\mu}{T}}-1} \qquad F_{eq} = \frac{1}{e^{\frac{p-\mu-\mu_q}{T}}+1} \qquad \bar{F}_{eq} = \frac{1}{e^{\frac{p-\mu+\mu_q}{T}}-1}
\end{eqnarray}
We compute the three parameters, $T$, $\mu$ and $\mu_q$ by solving the system of equations
\begin{eqnarray}
    \epsilon = \epsilon_{eq} \\
    n = n_{eq} \\
    n^q - n^{\bar{q}} = n^q_{eq} - n^{\bar{q}}_{eq}
\end{eqnarray}
There are a couple of scenarios to consider:
\begin{itemize}
    \item If there is quark/antiquark symmetry. In this case, $\mu_q=0$, and the last equation will be trivially satisfied, so we only need to take care of the first two.
    \item If the $C_{1\leftrightarrow2}$ kernel is active, then $\mu=0$ and the second equation is equivalent to the first one, so we don't need to solve it.
\end{itemize}

The three possible scenarios are contemplated in the code. At the initial time, the equilibrium parameters are computed by solving the system of equations with the Newton-Raphson method. This needs to be fed with some initialization parameters, which may change depending on the initial condition. The following time steps use as initialization parameters for the solver the results obtained in the previous time step. Since we do not expect abrupt changes in the distributions, the parameters will be very similar between close time steps, and the algorithm will just refine their values.

\subsection{The GPU algorithm}

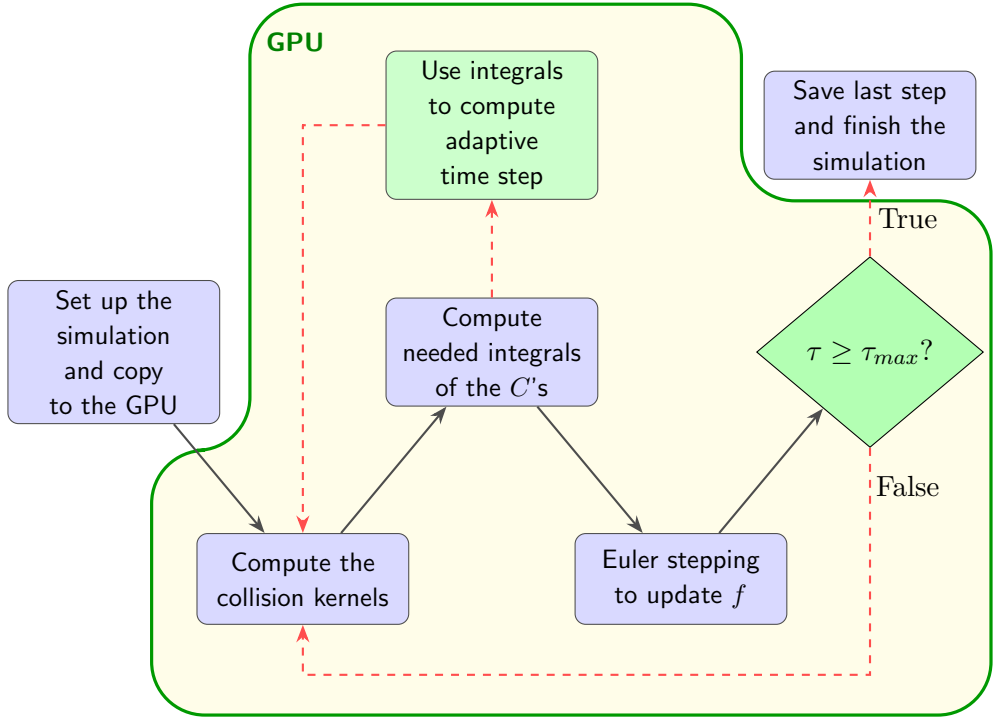
\begin{figure}
    \centering
\begin{tikzpicture}[
    box/.style={
        rectangle,
        rounded corners=4pt,
        draw=black!70,
        fill=blue!15,
        minimum width=2.8cm,
        minimum height=1.2cm,
        text width=2.4cm,
        text centered,
        font=\sffamily\small
    },
    arr/.style={
        ->, >=Stealth,
        thick,
        draw=black!70
    },
    feedback/.style={
        ->, >=Stealth,
        thick,
        draw=red!70,
        dashed
    },
    side/.style={
        rectangle,
        rounded corners=4pt,
        draw=black!70,
        fill=green!20,
        minimum width=2.8cm,
        minimum height=1.2cm,
        text width=2.4cm,
        text centered,
        font=\sffamily\small
    }, 
]

\draw[
    decorate,
    draw=green!60!black,
    fill=yellow!10,
    line width=1.2pt,
    rounded corners=20pt
]
  (0.5, -4.8)
  -- (11.6, -4.8)
  -- (11.6,  2.)
  -- (8.3,  2.)
  -- (8.3,  4.6)
  -- (1.8,  4.6) 
  -- (1.8, -1.3) 
  -- (0.5, -1.3) 
  -- cycle;

\node[font=\sffamily\bfseries\small, text=green!50!black]
    at (2.4, 4.1) {GPU};

\node[box] (n1) at (0,    0)  {Set up the simulation and copy to the GPU};
\node[box] (n2) at (2.5,   -3)  {Compute the collision kernels};
\node[box] (n3) at (5,    0)  {Compute needed integrals of the $C$'s};
\node[box] (n4) at (7.5,  -3)  {Euler stepping to update $f$};
\node[decision] (n5) at (10, 0) {$\tau\geq \tau_{max} ?$};
\node[box] (n6) at (10, 3) {Save last step and finish the simulation};

\draw[arr] (n1) -- (n2);
\draw[arr] (n2) -- (n3);
\draw[arr] (n3) -- (n4);
\draw[arr] (n4) -- (n5);
 
\node[side] (side) at (5, 3) {Use integrals to compute adaptive time step};
\draw[feedback] (n3) -- (side);
\draw[feedback] (side.west) -- ++ (-1.1, 0) -| (n2.north);
 
\draw[feedback]
    (n5.south) 
    -- ++(0, -3)
    -| (n2.south);

\node[style] at ($(n5.south) + (0.5, -0.5)$) {False};

\draw[feedback]
    (n5.north) -- (n6.south);
\node[style] at ($(n5.north) + (0.5, 0.5)$) {True};

\end{tikzpicture}

    \caption{Flux diagram of the BEDA solver algorithm}
    \label{fig:fluxdiagram}
\end{figure}

The calculation of the three elements of Eq.~\eqref{eq:BE_num} can be highly parallelized for each of the time steps. Because of this, we have implemented the numerical solver to run on GPUs. The main structure of the algorithm is sketched in Fig.~\ref{fig:fluxdiagram}, and we comment on the key aspects in the following. 

Once the initial condition has been set up and copied to the GPU, the collision kernels are computed. The algorithm to compute the $2\leftrightarrow2$ requires knowing the $1\leftrightarrow2$ as well as the expansion kernels. Thus, the last two are computed before the former. The computation of the collision kernels is parallelized over all the elements of the momentum grid, since each of them involves an independent calculation\footnote{This is not true for the case of the $2\leftrightarrow2$. In this case, the algorithm requires inverting a tridiagonal matrix. This happens for each of the steps described in Section~\ref{sec:C22}, but it only involves the inversion of the matrix for the current implicitly-integrated dimension. Thus, the parallelization is performed over the other 2 dimensions for which the matrix is, in principle, different.}. For the case of a $64\times 64\times 64$ grid, we are already exploiting as much as possible all the parallelization capability of a GPU ($\sim$ 10000 cores).

For each collision kernel, we compute a few integral moments that are used to determine the time step that will be used in the next time step\footnote{Because the $C_{2\leftrightarrow 2}$ collision kernel has been partially integrated with an implicit method, it is not possible to update the current time step.}. The set of integrals we use is
\begin{eqnarray}
    \delta I = \bigg\lbrace \int d^3 \mathbf{p} \frac{C}{p}~,~ \int d^3 \mathbf{p} C ~,~ \int d^3 \mathbf{p}pC~,~ \int d^3 \mathbf{p} \frac{C}{p^{3/2}}~,~ \int d^3 \mathbf{p} \frac{p_z^2}{p}C~\bigg\rbrace.
\end{eqnarray}
A similar set of integrals $I$ is defined by computing the same moments for the distribution function. With them, we can obtain the relative time variation for each of these quantities. Similarly to the adaptive time step described in~\cite{Du:2020zqg, Boguslavski:2024kbd}, we impose that, for each time step, the $\delta I /I$ ratio is smaller than a given quantity that we enforce to be a permille. Thus, our criterion to choose the next time step is given by
\begin{eqnarray}
    \Delta t = \left( \Delta t_{old}^3 \frac{0.001}{\max(|\delta I / I|)}\right)^{1/4}~.
\end{eqnarray}
In this expression, $\Delta t_{old}$ is the previous time step, and the power of 3 in it and the overall power of $1/4$ are meant to avoid an abrupt transition from the previous time step to the one we have just computed.

\begin{acknowledgments}
S.B.C. thanks A. Mazeliauskas for useful discussions. This work is supported by the European Research Council under project ERC-2018-ADG-835105 YoctoLHC; by Maria de Maeztu excellence unit grant CEX2023-001318-M and project PID2023-152762NB-I00 funded by MICIU/AEI/10.13039/501100011033; and by ERDF/EU. It has received funding from Xunta de Galicia (CIGUS Network of Research Centres). 
S.B.C. is supported by the DFG through the Emmy Noether Programme (project number 496831614) and CRC 1225
ISOQUANT (project number 27381115). B.W. acknowledges the support of the Ram\'{o}n y Cajal program with the Grant No. RYC2021-032271-I and the support of Xunta de Galicia under the ED431F 2023/10 project.
\end{acknowledgments}

\bibliographystyle{JHEP}
\bibliography{bulk}

\end{document}